\documentclass[12pt]{article}
\usepackage{amsmath,amssymb,graphicx,multirow,subfigure,oldgerm,euscript}
\usepackage[section]{placeins}

\begin{document}
\begin{titlepage}
\begin{flushright}
PCCF-RI-02-07 \\
ADP-02-94/T532  
\end{flushright}
\renewcommand{\thefootnote}{\fnsymbol{footnote}}
\vspace{-0.5em}
\begin{center}
{\LARGE{{  Direct $CP$ Violation in $B \rightarrow \pi^{+} \pi^{-} \pi$ \\
\vspace{0.6em}
 Determination of $\alpha$ without discrete ambiguity}}}
\end{center}
\vspace{-0.5em}
\begin{center}
\begin{large}
O. Leitner$^{1,2}$\footnote{oleitner@physics.adelaide.edu.au}, 
X.-H. Guo$^{1}$\footnote{xhguo@physics.adelaide.edu.au},  
A.W. Thomas${^1}$\footnote{athomas@physics.adelaide.edu.au} \\
\end{large}
\vspace{1.5em}
$^1$ Department of Physics and Mathematical Physics, and \\
Special Research Center for the Subatomic Structure of Matter, \\
University of Adelaide, Adelaide 5005, Australia \\
\vspace{0.5em}
$^2$ Laboratoire de Physique Corpusculaire, Universit\'e Blaise Pascal, \\
CNRS/IN2P3, 24 avenue des Landais, 63177 Aubi\`ere Cedex, France 
\end{center}
\vspace{2.5em}
\begin{abstract}
\vspace{1.0em}
Direct $CP$ violation in  the hadronic decays $\Bar{B}^{0} \rightarrow \pi^{+}\pi^{-} \pi^{0}$ is investigated near
 the peak  of  the $\rho^{0}$ taking  into account     the effect of $\rho - \omega$ mixing. 
Branching ratios  for  processes   $B^{\pm,0} \rightarrow \rho^{\pm,0}\pi^{\pm,0}$ and 
$B^{-} \rightarrow \omega \pi^{-}$ are calculated as well. 
We find that the $CP$ violating asymmetry is strongly dependent on the CKM matrix elements. For a fixed $N_{c}^{eff}$, 
the $CP$ violating asymmetry, $a_{CP}$, has a maximum of order $-40\%$ to $-70\%$ for
 $\Bar{B}^{0} \rightarrow \rho^{0}(\omega)
\pi^{0}$ 
when the invariant mass of the $\pi^{+}\pi^{-}$ pair is in the vicinity of the $\omega$ resonance.
 The sensitivity of the asymmetry  to $N_{c}^{eff}$ is small  in that case. Moreover, we find that in the range of 
  $N_{c}^{eff}$ which is allowed by    the most recent   experimental branching ratios  from the BABAR, BELLE 
and CLEO Collaborations, the sign of $\sin \delta$ is always positive. Thus, a measurement of direct $CP$
 violation in decays $\Bar{B}^{0}  \rightarrow \pi^{+}\pi^{-} \pi^{0}$ would remove the mod$(\pi)$ ambiguity
 in the determination of the $CP$ violating phase angle $\alpha$.
\end{abstract}
\vspace{7.5em}
%
%
%
%
\end{titlepage}
\newpage
%
%
%
\section{Introduction}\label{intro}
%
%
In the Standard Model,  $CP$ violating phenomena arise from a non-zero weak phase angle in
a complex matrix allowing flavour violation in the weak interaction:  the Cabbibo-Kobayashi-Maskawa (CKM) matrix.
Although the source of $CP$ violation has  not been well understood up to now,  physicists
are striving  to increase their  knowledge of  the mechanism. Many theoretical 
 studies~\cite{ref1,ref2} (within and beyond the Standard Model) and experimental investigations have been
conducted since the discovery of $CP$ violation  in neutral Kaon decays (1964). According to theoretical
predictions, large $CP$ violating effects may be expected in $B$ meson decays. In the past few years, several
facilities have started to collect events on $B$ decays and most of them refer to branching ratios. Generally, the main
theoretical uncertainties  apart from the    CKM matrix elements are the 
 hadronic matrix elements, where non-factorizable 
effects are involved.
As regards hadronic matrix elements and non-factorizable effects,  a new 
QCD factorization approach~\cite{ref3a}  has been proposed. 
This QCD factorization approach includes all radiative diagrams (gluon exchange) but  will not be the subject of 
this paper. For the CKM matrix elements,  uncertainties in the  parameters $\rho$ and $\eta$ 
have been reduced and this allows  us to predict $CP$ violating asymmetry in $B$ decays more accurately than before. This
will give us an excellent test for the Standard Model  and may lead to suggestions of new physics.
\newline
Direct  $CP$ violating asymmetries in $B$ decays  occur  through the interference of at least two amplitudes with
 different weak  phase $\phi$ {\em and} strong phase $\delta$. In order to extract  the weak phase  
 (which is determined by the CKM matrix elements),  one must know  the strong phase   $\delta$ and this 
is usually not well determined. In addition, in order to have a large signal, we have to appeal to  some
 phenomenological mechanism to obtain  a large $\delta$. The charge symmetry violating mixing between
 $\rho^{0}$ and $\omega$  can be extremely important in this regard. In particular, it can lead to   a 
 large $CP$ violation in $B$ decay such as $\Bar{B}^{0} \rightarrow \rho^{0}(\omega) \pi^{0} \rightarrow \pi^{+}
 \pi^{-}  \pi^{0}$, because   the strong phase  passes  through $90^{o}$ at the $\omega$ 
resonance~\cite{ref3,ref4,ref5}. 
\newline
We have collected all the recent  data for $b$ to $d$ transitions, but we shall focus on the   CLEO, BABAR and
 BELLE branching ratio results. 
We also shall use the latest values for CKM parameters, $A, \lambda, \rho$, and $\eta$. The aim of the present
work is to constrain the $CP$ violating calculation in $\Bar{B}^{0} \rightarrow \rho^{0}(\omega) 
\pi^{0}\rightarrow \pi^{+} \pi^{-} \pi^{0}$, including   $\rho-\omega$ mixing and  using the most recent 
experimental  data for the  branching ratios for $B \rightarrow \rho \pi$ decays.  In order to extract the 
strong phase $\delta$, we  use the naive factorization approach, in which the hadronic matrix elements of 
operators are saturated by vacuum intermediate states. Moreover, we approximate non-factorizable effects by 
introducing an effective number of colours, $N_{c}^{eff}$.
\newline
In this paper, we  investigate five  phenomenological models with different weak form factors and  determine
 the $CP$ violating asymmetry for $\Bar{B}^{0} \rightarrow \rho^{0}(\omega) \pi^{0} \rightarrow \pi^{+} 
\pi^{-}\pi^{0} $  in  these  models. We select models which are consistent with all the latest 
data and  determine the  allowed range for $N_{c}^{eff}$ ($1.09(1.11)< N_{c}^{eff}<1.68(1.80)$). Then,  we  study  
the sign of  $\sin \delta$ in this range of $N_{c}^{eff}$ in  all these models. We  also discuss the model  dependence 
of our results  in detail.
\newline
This paper is structured as follows. In Section 2, we introduce the effective 
 Hamiltonian based on  the Operator Product Expansion (OPE) including  Wilson coefficients.  We also present the
 formalism of
$\rho-\omega$ mixing and its application to the $CP$ violating asymmetry in decay processes. In Section 3, 
 the CKM matrix  and the relevant form factors are discussed. In Section 4, we 
present  numerical results
for the $CP$ violating  asymmetry in $\Bar{B}^{0}  \rightarrow \pi^{+}\pi^{-} \pi^{0}$  which is followed by 
discussion of  these results. In Section 5, branching ratios for decays such as $B^{\pm,0} \rightarrow 
\rho^{\pm,0} \pi^{\pm,0}$ and $B^{-} \rightarrow \omega \pi^{-}$  are investigated.
 From  the CLEO, BABAR and BELLE experimental data for these branching ratios, we extract the range of $N_{c}^{eff}$ 
allowed in these processes and the results are also discussed. In the final section, we summarize our results.  
Comments on form factors,  CKM matrix parameter values, $\rho, \eta$, and conclusions are also given in this section. 
%
%
\section{$\boldsymbol{CP}$ violation in $\boldsymbol{\bar{B}^{0}\rightarrow  \rho^{0}  \pi^{0}  \rightarrow \pi^{+} \pi^{-} \pi^{0}}$}
\label{part1}
%
%
\subsection{Effective  theory}\label{part1.1}
%
%
In any phenomenological treatment of the weak decays of hadrons, the starting point is the weak effective
Hamiltonian at low energy~\cite{ref13} from which, the decay amplitude can  be expressed as  follows, 
\begin{multline}\label{eq1.3}
A(B \rightarrow P V) =
\frac {G_{F}}{\sqrt 2} \biggl[  V_{ub}V_{ud}^{\ast}\bigl( C_{1}\langle P V | O_{1}^{u}| B \rangle + 
C_{2}\langle P V |O_{2}^{u}| B \rangle \bigr)  \\
-  V_{tb}V_{td}^{\ast} \sum_{i=3}^{10} C_{i}\langle P V |O_{i}| B \rangle \biggr]+ h.c.\ ,
\end{multline}
where $\langle P V |O_{i}| B \rangle$ are the hadronic matrix elements.  They  describe the transition between initial
 and final states with the operator renormalized at scale  $\mu$  and include, up to now, the main  
uncertainties in the calculation since they involve   non-perturbative effects. $G_{F}$ is the Fermi 
constant, $V_{CKM}$ is the CKM matrix element, 
$C_{i}(\mu)$  are the Wilson coefficients, $O_i(\mu)$ are the operators from 
OPE~\cite{ref6}. The operators $O_{i}$,  the local operators
 which govern weak  decays  can be written  as,
\begin{align}
O_{1}^{u}& = \bar{q}_{\alpha} \gamma_{\mu}(1-\gamma{_5})u_{\beta}\bar{u}_{\beta} \gamma^{\mu}(1-\gamma{_5})
b_{\alpha}\ , & O_{2}^{u}& = \bar{q} \gamma_{\mu}(1-\gamma{_5})u\bar{u} \gamma^{\mu}(1-\gamma{_5})b\ ,  \nonumber \\
O_{3}& = \bar{q} \gamma_{\mu}(1-\gamma{_5})b \sum_{q\prime}\bar{q}^{\prime}\gamma^{\mu}(1-\gamma{_5})
q^{\prime}\ , & O_{4}& =\bar{q}_{\alpha} \gamma_{\mu}(1-\gamma{_5})b_{\beta} 
\sum_{q\prime}\bar{q}^{\prime}_{\beta}\gamma^{\mu}(1-\gamma{_5})q^{\prime}_{\alpha}\ , \nonumber \\
O_{5}& =\bar{q} \gamma_{\mu}(1-\gamma{_5})b \sum_{q'}\bar{q}^
{\prime}\gamma^{\mu}(1+\gamma{_5})q^{\prime}\ , & O_{6}& =\bar{q}_{\alpha} \gamma_{\mu}(1-\gamma{_5})b_{\beta} 
\sum_{q'}\bar{q}^{\prime}_{\beta}\gamma^{\mu}(1+\gamma{_5})q^{\prime}_{\alpha}\ ,  \nonumber \\ 
O_{7}& =\frac{3}{2}\bar{q} \gamma_{\mu}(1-\gamma{_5})b \sum_{q'}e_{q^{\prime}}
\bar{q}^{\prime} \gamma^{\mu}(1+\gamma{_5})q^{\prime}\ , & O_{8}& =\frac{3}{2}\bar{q}_{\alpha} 
\gamma_{\mu}(1-\gamma{_5})b_{\beta} 
\sum_{q'}e_{q^{\prime}}\bar{q}^{\prime}_{\beta}\gamma^{\mu}(1+\gamma{_5})q^{\prime}_{\alpha}\ , \nonumber \\
O_{9}& =\frac{3}{2}\bar{q} \gamma_{\mu}(1-\gamma{_5})b \sum_{q'}e_{q^{\prime}}
\bar{q}^{\prime} \gamma^{\mu}(1-\gamma{_5})q^{\prime}\ , & O_{10}& =\frac{3}{2}\bar{q}_{\alpha}
 \gamma_{\mu}(1-\gamma{_5})b_{\beta} \sum_{q'}e_{q^{\prime}}\bar{q}^{\prime}_{\beta}
\gamma^{\mu}(1-\gamma{_5})q^{\prime}_{\alpha}\ ,
\end{align}
where $q^{\prime}= u,d,s,c$ and $e_{q^{\prime}}$ denotes its  electric  charge. As regards the  Wilson 
coefficients~\cite{ref9, ref10, ref11, ref12}, they represent the physical contributions from scales higher 
than $\mu$. Since  QCD has the property of  asymptotic freedom, they can 
be calculated in perturbation theory.  Usually, the scale $\mu$ is chosen to be of   
  order $O(m_{b})$ for $B$ decays and  Wilson coefficients have been  calculated to the next-to-leading order (NLO).
For more details see Ref.~\cite{ref14}.

%
\subsection{$\boldsymbol{\rho -\omega}$ Mixing}\label{part2.1}

Let $A$ be the amplitude for the decay $B \rightarrow \rho^{0}  \pi \rightarrow  \pi^{+}  \pi^{-} \pi $ 
 then one has,
\begin{equation}\label{eq1.15}
A=\langle  \pi  \pi^{-} \pi^{+}|H^{T}|B \rangle + \langle  \pi  \pi^{-} \pi^{+}|H^{P}|B  \rangle\ ,
\end{equation}
with $H^{T}$ and $H^{P}$ being the Hamiltonians for the tree and penguin operators. We can 
define the relative magnitude and phases between these two contributions  as follows,
\begin{align}\label{eq1.16}
A &= \langle  \pi  \pi^{-} \pi^{+}|H^{T}| B \rangle [ 1+re^{i\delta}e^{i\phi}]\ , \nonumber \\   
\bar {A} &= \langle \bar{\pi}  \pi^{+}  \pi^{-}|H^{T}|\bar {B} \rangle [ 1+re^{i\delta}e^{-i\phi}]\ ,  
\end{align}
where $\delta$ and $\phi$ are the strong and weak phases, respectively. The phase $\phi$ arises from the 
appropriate combination of CKM matrix elements, and, assuming top quark dominance,
$ \phi={\rm arg}[(V_{tb}V_{td}^{\star})/(V_{ub}V_{ud}^{\star})]$. 
As a result, $\sin \phi$ is equal 
to $\sin \alpha$, with $\alpha$ defined in the standard way~\cite{ref16}. The parameter, $r$, is the 
absolute value of the ratio of tree and penguin amplitudes:
\begin{equation}\label{eq1.17}
r \equiv \left| \frac{\langle \rho^{0} \pi|H^{P}|B \rangle}{\langle\rho^{0}
\pi |H^{T}|B \rangle} \right|.
\end{equation}
In order to obtain a  large signal for direct $CP$ violation, we need some mechanism to make both 
$\sin\delta$  and  $r$ large. We stress that   $\rho-\omega$ mixing~\cite{refa2} has the dual advantages that the
 strong 
phase difference is large (passing through $90^{o}$ at the $\omega$ resonance) and well known~\cite{ref4, ref5}.
With this mechanism, to first  order in  isospin violation, we have the following results when the invariant 
mass of $\pi^{+}\pi^{-}$ is near the $\omega$ resonance mass,
\begin{align}\label{eq1.18}
\langle \pi  \pi^{-} \pi^{+}|H^{T}|B  \rangle & = \frac{g_{\rho}}{s_{\rho}s_{\omega}}
 \tilde{\Pi}_{\rho \omega}t_{\omega} +\frac{g_{\rho}}{s_{\rho}}t_{\rho}\ , \nonumber \\
\langle  \pi \pi^{-} \pi^{+}|H^{P}|B  \rangle & = \frac{g_{\rho}}{s_{\rho}s_{\omega}} 
\tilde{\Pi}_{\rho \omega}p_{\omega} +\frac{g_{\rho}}{s_{\rho}}p_{\rho}\ .
\end{align}
Here $t_{V} (V=\rho \;{\rm  or} \; \omega) $ is the tree amplitude and $p_{V}$ the penguin amplitude for 
 producing a vector meson, V, $g_{\rho}$ is the coupling for $\rho^{0} \rightarrow \pi^{+}\pi^{-}$, 
$\tilde{\Pi}_{\rho \omega}$ is the effective $\rho-\omega$ mixing amplitude, and $s_{V}$  is  from the inverse 
 propagator of the vector meson V,
\begin{equation}\label{eq1.19}
s_{V}=s-m_{V}^{2}+im_{V}\Gamma_{V}\ , 
\end{equation}
with $\sqrt s$ being the invariant mass of the $\pi^{+}\pi^{-}$ pair. 
We stress that the direct coupling $ \omega \rightarrow \pi^{+} \pi^{-} $ is effectively absorbed into 
$\tilde{\Pi}_{\rho \omega}$~\cite{ref17},  leading  to the explicit $s$ dependence of $\tilde{\Pi}_{\rho \omega}$. 
Making the expansion  $\tilde{\Pi}_{\rho \omega}(s)=\tilde{\Pi}_{\rho \omega}(m_{\omega}^{2})+(s-m_{w}^{2}) 
\tilde{\Pi}_{\rho \omega}^{\prime}(m_{\omega}^{2})$, the  $\rho-\omega$ mixing parameters were determined in 
the fit of Gardner and O'Connell~\cite{ref18}: $\Re e \; 
\tilde{\Pi}_{\rho \omega}(m_{\omega}^{2})=-3500 \pm 300 \; {\rm MeV}^{2}, \;\;\; \Im m \; \tilde{\Pi}_{\rho \omega}
(m_{\omega}^{2})= -300 \pm 300 \; {\rm MeV}^{2}$ and  $\tilde{\Pi}_{\rho \omega}^{\prime}
(m_{\omega}^{2})=0.03 \pm 0.04$. In practice, the effect of the derivative term is negligible.
%
%
From Eqs.~(\ref{eq1.15},~\ref{eq1.16},~\ref{eq1.18}) one has,
\begin{equation}\label{eq1.20}
 re^{i \delta} e^{i \phi}= \frac{ \tilde {\Pi}_{\rho \omega}p_{\omega}+s_{\omega}p_{\rho}}{\tilde 
{\Pi}_{\rho \omega} t_{\omega} + s_{\omega}t_{\rho}}\ . 
\end{equation}
Defining,
\vspace{-1em}
\begin{center}
\begin{equation}\label{eq1.21}
\frac{p_{\omega}}{t_{\rho}} \equiv r^{\prime}e^{i(\delta_{q}+\phi)}\ , \;\;\;\;
\frac{t_{\omega}}{t_{\rho}} \equiv \alpha e^{i \delta_{\alpha}}\ , \;\;\;\;
\frac{p_{\rho}}{p_{\omega}} \equiv \beta e^{i \delta_{\beta}}\ , 
\end{equation}
\end{center}
where $ \delta_{\alpha}, \delta_{\beta}$, and $ \delta_{q}$ are strong phases (absorptive part).
Substituting Eq.~(\ref{eq1.21}) into  Eq.~(\ref{eq1.20}),   one finds,
\begin{equation}\label{eq1.22}
re^{i\delta}=r^{\prime}e^{i\delta_{q}} \frac{\tilde{\Pi}_{\rho \omega}+ \beta e^{i \delta_{\beta}} 
s_{\omega}}{s_{\omega}+\tilde{\Pi}_{\rho \omega} \alpha e^{i \delta_{\alpha}}}\ .
\end{equation}
$\alpha e^{i \delta_{\alpha}}$, $\beta e^{i \delta_{\beta}}$, and $r^{\prime}e^{i \delta_{q}}$ will be
 calculated later. In order to get the $CP$ violating asymmetry $a_{CP}$, 
$\sin\phi$ and $\cos\phi$ are needed, where $\phi$ is determined by the CKM matrix elements. 
In the Wolfenstein parametrization, the weak phase comes from 
$[V_{tb} V^{\star}_{td}/V_{ub} V^{\star}_{ud}]$ and one has for the decay $B \rightarrow \rho(\omega) \pi$,
\begin{align}\label{eq1.27}
\sin\phi & = \frac{\eta}{\sqrt {[\rho(1-\rho)-\eta^{2}]^{2}+\eta^{2}}}\ , \nonumber \\
\cos\phi & = \frac{\rho(1-\rho)-\eta^{2}}{\sqrt {[\rho(1-\rho)-\eta^{2}]^{2}+\eta^{2}}}\ .
\end{align}
The values used for $\rho$ and $\eta$ will be discussed in Section~\ref{part3.1}.
With the decay amplitude given in Eq.~(\ref{eq1.3}),  we are ready to evaluate 
the matrix elements  for $B^{\pm,0}\rightarrow 
\rho^{0}(\omega) \pi^{\pm,0}$.  In the factorization 
approximation~\cite{ref19}, either  $\rho^{0}(\omega)$ or  
$\pi^{\pm,0}$ is generated by one current which has the appropriate  quantum numbers in the Hamiltonian.
For these  decay processes, two kinds of matrix element products are involved after factorization; schematically 
(i.e. omitting Dirac matrices and colour labels) one has  $\langle \rho^{0}(\omega)|(\bar{u}u)|0\rangle 
\langle \pi^{\pm,0}|(\bar{u} b)|B^{\pm,0}\rangle $ and $ \langle \pi^{\pm,0}|
(\bar{q}_{1}q_{2})|0\rangle \langle\rho^{0}
(\omega)|(\bar{u}b)|B^{\pm,0}\rangle$ with $q_{i}(i=1,2) = u, d$. We will calculate 
them in some phenomenological quark models. 
\newline
The matrix elements for $B \rightarrow X$ and  $B \rightarrow X^{\star}$ (where X and $ X^{\star}$ 
denote pseudoscalar and vector mesons, respectively) can be decomposed as follows~\cite{ref20},
\begin{equation}\label{eq1.28}
\langle X|J_{\mu}|B \rangle =\left( p_{B} + p_{X}- \frac{m_{B}^{2}-m_{X}^{2}}{k^{2}}k \right)_{\mu} 
F_{1}(k^{2})+\frac{m_{B}^{2}-m_{X}^{2}}{k^{2}}k_{\mu}F_{0}(k^{2})\ ,
\end{equation}
and 
\begin{multline}\label{eq1.29}
\langle X^{\star}|J_{\mu}|B \rangle=\frac{2}{m_{B}+m_{X^{\star}}} \epsilon_{\mu \nu \rho \sigma} 
\epsilon^{\star \nu} p_{B}^{\rho} p_{X^{\star}}^{\sigma}  V(k^{2}) +i \Biggl\{ \epsilon_{\mu}^{\star}(m_{B}
+m_{X^{\star}})A_{1}(k^{2})  \\
- \frac{\epsilon^{\star}  \cdot k}{m_{B}+m_{X^{\star}}} (P_{B}+P_{X^{\star}})_{\mu}A_{2}(k^{2})- \frac 
{\epsilon^{\star}  \cdot k}{k^{2}}2m_{X^{\star}} \cdot k_{\mu}A_{3}(k^{2})
\Biggr\}  \\
+i \frac{\epsilon^{\star} \cdot k}{ k^{2}}2m_{X^{\star}} \cdot k_{\mu}A_{0}(k^{2})\ ,
\end{multline}
where  $J_{\mu}(=\bar{q}\gamma^{\mu}(1-\gamma_{5})b)$     
 is the weak current   with  $ q=u,d$,  $k=p_{B}-p_{X(X^{\star})}$, and  $\epsilon_{\mu}$ is the polarization 
vector of $X^{\star}$.  $F_{0}$ and $ F_{1}$ are the form factors related to the transition $0^{-} 
\rightarrow 0^{-}$ and  $A_{0}, A_{1}, A_{2}, A_{3}$, and $ V$ are the form factors which describe the 
transition $0^{-} \rightarrow 1^{-}$.
Finally, in order to cancel the poles at $q^{2}=0$, the form factors must respect the constraints:
\begin{equation}\label{eq1.30}
F_{1}(0) = F_{0}(0), \;\;  A_{3}(0) = A_{0}(0)\ .
\end{equation}
They also  satisfy the following relations: 
\begin{gather}\label{eq1.31}
A_{3}(k^{2})  = \frac {m_{B}+m_{X^{\star}}}{2m_{X^{\star}}}A_{1}(k^{2})- \frac {m_{B}-m_{X^{\star}}}
{2m_{X^{\star}}}A_{2}(k^{2})\ . 
\end{gather}

\noindent  By using the decomposition in 
Eqs.~(\ref{eq1.28},~\ref{eq1.29}), one obtains  the following tree operator contribution for the process $\Bar{B}^{0} 
\rightarrow \rho^{0}(\omega)
\pi^{0}$:
\begin{equation}\label{eq1.32}
t_{\rho}=  m_{B}|\vec{p}_{\rho}|  \biggl[ (C_{1}^{\prime}+\frac {1}{N_{c}^{eff}}C_{2}^{\prime}) \biggr]
\biggl( f_{\rho}F_{1}(m_{\rho}^{2})+   f_{\pi}A_{0}(m_{\pi}^{2}) \biggr) \ ,
\end{equation}
where $f_{\rho}$ and $f_{\pi}$ are the decay constants of  $\rho$ and $\pi$, respectively, and $C_{i}^{\prime}$ are
 the Wilson coefficients with values listed in Table~\ref{tab2}.
We find $t_{\omega} \neq t_{\rho}$, so that
\begin{equation}\label{eq1.33}
\alpha e^{i \delta_{\alpha}}=\frac{-f_{\rho}F_{1}(m_{\rho}^{2})+ f_{\pi}A_{0}(m_{\pi}^{2})}
{f_{\rho}F_{1}(m_{\rho}^{2})+ f_{\pi}A_{0}(m_{\pi}^{2})}\ . 
\end{equation}
After calculating the penguin operator contributions, one has,
\begin{equation}\label{eq1.34}
r^{\prime}e^{i \delta_{q} }=-
\frac{p_{\omega}}{(C_{1}^{\prime}+\frac {1}{N_{c}^{eff}}C_{2}^{\prime}) 
\bigl( f_{\rho}F_{1}(m_{\rho}^{2})+   f_{\pi}A_{0}(m_{\pi}^{2})\bigr)}\left| 
\frac{V_{tb}V_{td}^{\star}}{V_{ub}V_{ud}^{\star}}\right|\ , 
\end{equation}
and
\begin{multline}\label{eq1.35}
\beta e^{i \delta_{\beta}}=
\frac{ m_{B}| \vec{p}_{\rho}|}{p_{\omega}} \Bigg\{ -(C_{4}^{\prime}+\frac{1}{N_{c}^{eff}}C_{3}^{\prime})
[f_{\rho}F_{1}(m_{\rho}^{2})+f_{\pi}A_{0}(m_{\pi}^{2})] \\
\hspace{5em} -\frac{3}{2}[(C_{7}^{\prime}+\frac{1}{N_{c}^{eff}}C_{8}^{\prime})-(C_{9}^{\prime}+
\frac{1}{N_{c}^{eff}}C_{10}^
{\prime})] f_{\pi}A_{0}(m_{\pi}^{2}) \\
\hspace{5em} +\frac{3}{2}[(C_{7}^{\prime}+\frac{1}{N_{c}^{eff}}C_{8}^{\prime})+(C_{9}^{\prime}+
\frac{1}{N_{c}^{eff}}C_{10}^
{\prime})]  f_{\rho}F_{1}(m_{\rho}^{2})  \\
+[(C_{6}^{\prime}+\frac{1}{N_{c}^{eff}}C_{5}^{\prime})-\frac{1}{2}(C_{8}^{\prime}+
\frac{1}{N_{c}^{eff}}C_{7}^{\prime})] 
\left[ \frac{2m_{\pi}^{2}f_{\pi}A_{0}(m_{\pi}^{2})}{(m_{d}+m_{d})(m_{b}+m_{d})}\right]   \\
+ \frac{1}{2}(C_{10}^{\prime}+\frac{1}{N_{c}^{eff}}C_{9}^{\prime})[f_{\rho}F_{1}(m_{\rho}^{2})+f_{\pi}
A_{0}(m_{\pi}^{2})]\Bigg\}\ ,
\end{multline}
\vspace{-1em}
where  $\vec{p}_{\rho}$ is the
c.m. momentum of the decay process.
In  Eqs.~(\ref{eq1.34}, \ref{eq1.35}),  $p_{\omega}$  is written as,
\begin{multline}\label{eq1.36}
p_{\omega}= m_{B}|\vec{p}_{\rho}| \Bigg\{ 
-2 \left[ (C_{3}^{\prime}+\frac {1}{N_{c}^{eff}}C_{4}^{\prime})+
(C_{5}^{\prime}+\frac {1}{N_{c}^{eff}}C_{6}^{\prime})\right]f_{\rho}F_{1}(m_{\rho}^{2}) \\
\hspace{5em} - \frac{1}{2}\left[(C_{7}^{\prime}+\frac {1}{N_{c}^{eff}}C_{8}^{\prime})+(C_{9}^{\prime}+
\frac {1}{N_{c}^{eff}}
C_{10}^{\prime})\right] f_{\rho}F_{1}(m_{\rho}^{2})  \\
-  \left[\frac{1}{2}(C_{8}^{\prime}+\frac {1}{N_{c}^{eff}}C_{7}^{\prime})-(C_{6}^{\prime}+\frac {1}{N_{c}^{eff}}C_{5}^{\prime})
 \right] \left[  \frac{2 m_{\pi}^{2}f_{\pi}A_{0}(m_{\pi}^{2})}{(m_{d}+m_{d})(m_{b}+m_{d})}\right]  \\ 
\hspace{9.5em} -  (C_{4}^{\prime}+\frac {1}{N_{c}^{eff}}C_{3}^{\prime}) \left[ f_{\pi}A_{0}(m_{\pi}^{2})+f_{\rho}F_{1}
(m_{\rho}^{2})\right]  \\
+  \frac{1}{2} (C_{10}^{\prime}+\frac {1}{N_{c}^{eff}}C_{9}^{\prime}) \left[ f_{\pi}A_{0}(m_{\pi}^{2})+\frac{1}{2}
f_{\rho}F_{1}(m_{\rho}^{2})\right] \Bigg\}\ ,    
\end{multline}
and the CKM amplitude entering  the $b \rightarrow d$ transition is, 
\begin{equation}\label{eq1.37}
\left| \frac{V_{tb}V_{td}^{\star}}{V_{ub}V_{ud}^{\star}}\right|
=\frac{\sqrt{{(1-\rho)}^{2}+{\eta}^{2}}}
{(1- {\lambda}^{2}/2)\sqrt{{\rho}^{2}+{\eta}^{2}}}=\left( 1- \frac{\lambda^{2}}{2} \right)^{-1} \left| 
\frac{\sin \gamma}{\sin \beta} \right|\ ,
\end{equation}
with $\beta$ and $\gamma$ defined  in the unitarity triangle as usual.
%
\section{Numerical inputs}\label{part3}
%
\subsection{CKM values and quark masses}\label{part3.1}
%
In our numerical calculations  we have several parameters:  $N_{c}^{eff}$ and the CKM matrix elements
in the Wolfenstein parametrization. The CKM matrix, which should be determined from
 experimental data,  is expressed  in terms  of the Wolfenstein parameters, $ A,\; \lambda,\; 
\rho$, and $ \eta $~\cite{ref15}.
Here we shall use  the latest  values~\cite{ref26}  which have been extracted from
charmless semileptonic $B$ decays ($|V_{ub}|$),  charmed semileptonic $B$ decays  ($|V_{cb}|$),
$s$ and $d$ mass oscillations 
and $CP$ violation in the kaon system $(\rho, \eta$):
\begin{equation}\label{eq1.45}
\lambda=0.2237\ , \;\; A=0.8113\ , \;\;  0.190 < \rho < 0.268\ , \;\; 0.284< \eta <0.366\ .
\end{equation}
These values  respect the unitarity triangle as well.
The running quark masses are used  in order to calculate the 
matrix elements of penguin operators. The quark mass is  taken at the scale $\mu \simeq m_{b}$ in  
$B$ decays. Therefore one has~\cite{ref27},
\vspace{-0.5em}
\begin{align}\label{eq1.46}
m_{u}(\mu=m_{b})& = 2.3 \;{\rm MeV}\ ,&  m_{b}(\mu=m_{b})& = 4.9 \;{\rm GeV}\ ,& 
m_{d}(\mu=m_{b})& = 4.6 \;{\rm MeV}\ , 
\end{align}
which corresponds to $m_{s}(\mu= 1\;{\rm GeV}) = 140 \;{\rm MeV}$. As regards meson  masses,
we shall use the following values~\cite{ref16}:
\begin{align}\label{eq1.47}
 m_{B^{0}}&   = 5.279 \; {\rm GeV}\ , & m_{\pi^{\pm}}& = 0.139 \;{\rm GeV}\ , & m_{\pi^{0}} & = 0.135 \;{\rm GeV}\ ,
\nonumber \\
 m_{\rho^{0}}& = 0.769  \;{\rm GeV}\ , &  m_{\omega}& = 0.782 \;{\rm GeV}\ .  
\end{align}
%
%
\subsection{Form factors and decay constants}\label{part3.3}
%
%
The form factors  $F_{i}(k^{2})$ and $A_{j}(k^{2})$  depend on the inner structure of  
hadrons. In order to gauge the model dependence of the results, we will adopt three different theoretical 
approaches. The first  was proposed  by Bauer, Stech,
and Wirbel~\cite{ref20} (BSW model). They used the overlap integrals of wave functions in order to evaluate 
the meson-meson matrix elements of 
the corresponding  current. 
The second approach was  developed by Guo and Huang (GH model)~\cite{ref22}. They modified the BSW model by using some 
wave functions 
described in the light-cone framework.
The last model was given  by Ball~\cite{ref23, ref24}. In this 
case, the form factors are  calculated from QCD sum rules on the light-cone and  leading twist contributions,
radiative corrections and $SU(3)$-breaking effects are included. The explicit $k^{2}$ dependence of 
the form factors is~\cite{ref20,ref22},
\begin{gather}\label{eq1.43}
F_{1}(k^{2})=\frac{h_{1}}{\left( 1-\frac{k^{2}}{m_{1}^{2}} \right)^{n}}\ , \;\;\;  \;\;\; 
A_{0}(k^{2})=\frac{h_{A_{0}}}{\left( 1-\frac{k^{2}}{m_{A_{0}}^{2}}\right)^{n}}\ , \nonumber \\
{\rm and}~\cite{ref23,ref24,ref25}      \hspace{+31.5em}       \nonumber \\
F_{1}(k^{2})=\frac{h_{1}}{1-d_{1}\frac{k^{2}}{m_{B}^{2}}+b_{1}\left( \frac{k^{2}}{m_{B}^{2}}\right)^{2}}\ ,
 \;\;\;  \;\;\; 
A_{0}(k^{2})=\frac{h_{A_{0}}}{1-d_{0}\frac{k^{2}}{m_{B}^{2}}+b_{0}\left( \frac{k^{2}}{m_{B}^{2}}\right)^{2}}\ ,
\end{gather} 
where $n=1,2$, and  $m_{A_{0}}$ and $m_{1}$ are the pole masses associated with   the transition current. $h_{1}$ 
and $h_{A_{0}}$
are the values of the corresponding form factors at $q^{2}=0$, and $d_{i},b_{i}$ ($i=0,1$) are  parameters in
 the  model of Ball.
In  Table~\ref{tab5} we list  the relevant form factor values at zero 
momentum transfer~\cite{ref20,ref22,ref23,ref24,ref28}  for 
$B \rightarrow \pi$ and  $B \rightarrow \rho$ transitions.
The different models are defined as follows: models (1) and (3) are the BSW models where the $q^{2}$ dependence of the 
form
factors is described  by a single and a double-pole ansatz, respectively. Models (2) and (4) 
 are the GH model with the 
same momentum dependence as  models (1) and (3). Finally,  model (5) refers to the Ball model.
We define  the decay constants for  pseudo-scalar ($f_{P}$) and vector ($f_{V}$) mesons  as usual by,
\begin{align}\label{eq1.48}
\langle P(q) | \bar{q}_{1} \gamma_{\mu} \gamma_{5} q_{2}| 0 \rangle & = -i f_{P} q_{\mu}\ , \nonumber \\
\sqrt{2} \langle V(q) | \bar{q}_{1} \gamma_{\mu} q_{2} | 0 \rangle & = f_{V} m_{V} \epsilon_{V}\ ,
\end{align}
with $q_{\mu}$ being the momentum of  the pseudo-scalar meson, and $m_{V}$ and $\epsilon_{V}$ being the mass
and polarization vector of the vector meson, respectively.
In our calculations we take~\cite{ref16}:
\begin{gather}\label{eq1.49}
   f_{\pi} = 132 \; {\rm MeV}\ ,  \; 
f_{\rho} \simeq f_{\omega}  = 221 \; {\rm MeV}\ . 
\end{gather}
In practise the $\rho$ and $\omega$ decay constants are very close, and as a   simplification (with little
 effect on the  results), we chose $f_{\rho} = f_{\omega}$.
%
%
\section{Results and discussions}\label{part4}
\begin{figure}
\centering\includegraphics[height=7cm,clip=true]{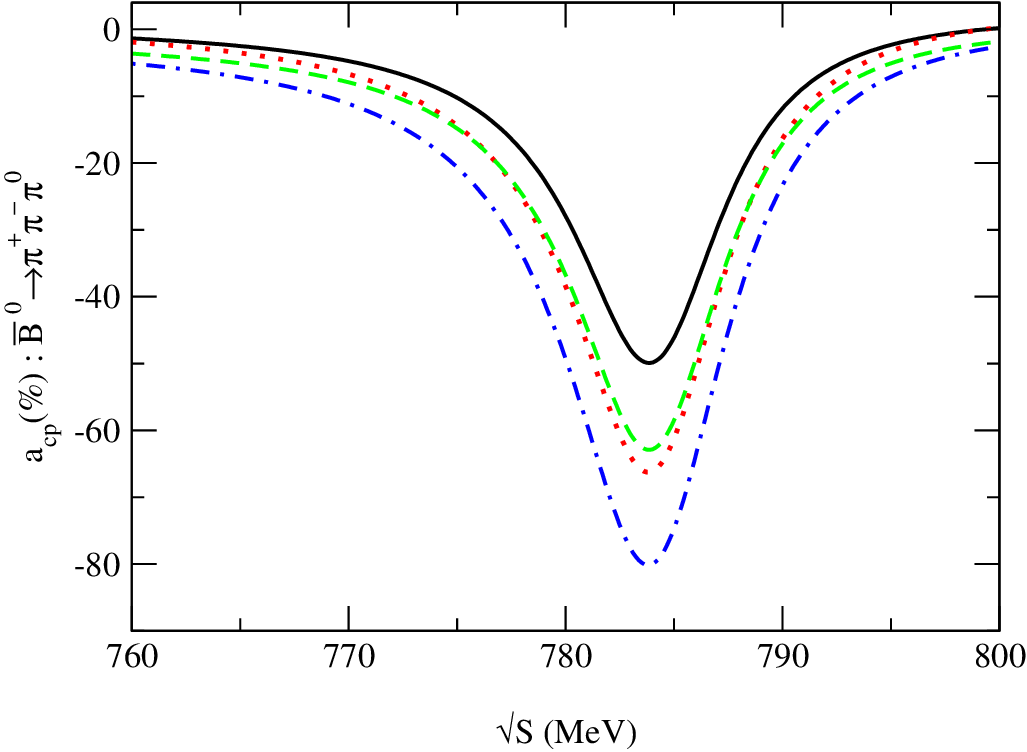}
\caption{$CP$ violating asymmetry, $a_{CP}$,  for $\bar{B}^{0} \rightarrow 
\pi^{+} \pi^{-} \pi^{0}$, for $k^{2}/m_{b}^{2}=0.3$, $N_{c}^{eff}=1.09(1.68)$  and limiting values of
 the CKM matrix elements for   model $(1)$: solid line (dotted line) for $N_{c}^{eff}=1.09$ and max(min) 
CKM matrix elements.  Dashed line (dot-dashed line) for $N_{c}^{eff}=1.68$ and max(min) CKM matrix elements.}
\label{fig6}
\end{figure}

\begin{figure}
\centering\includegraphics[height=7cm,clip=true]{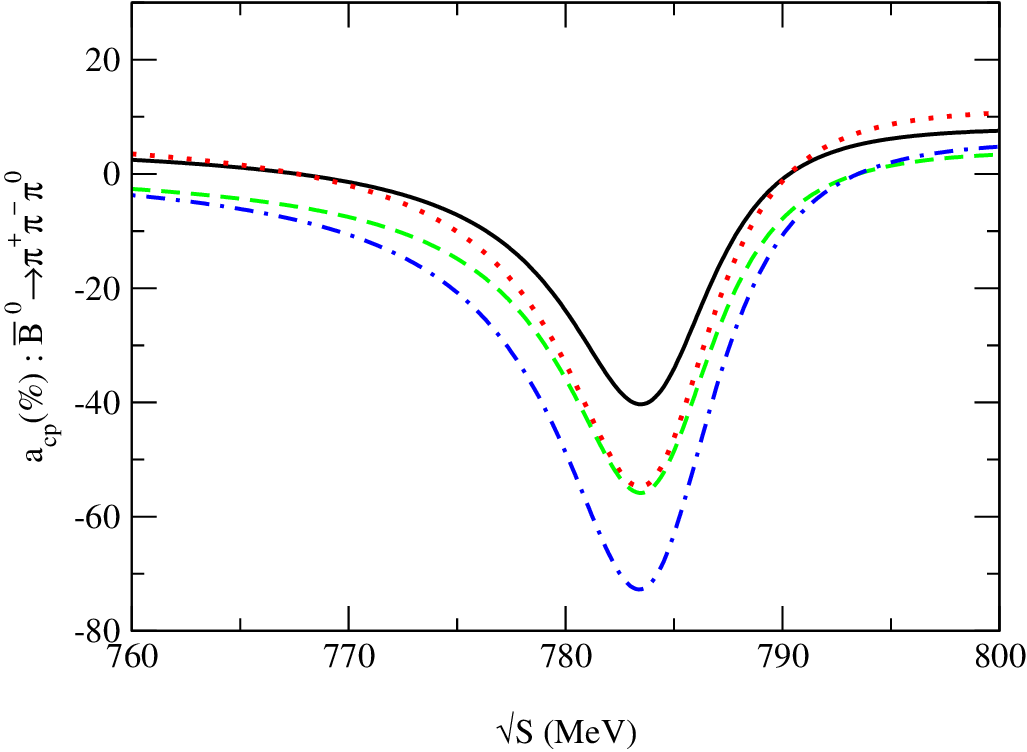}
\caption{$CP$ violating asymmetry, $a_{CP}$, for $\bar{B}^{0} \rightarrow 
\pi^{+} \pi^{-} \pi^{0}$, for $k^{2}/m_{b}^{2}=0.5$, $ N_{c}^{eff}=1.11(1.80)$  and limiting values of the 
 CKM matrix
 elements for  model $(1)$: solid line (dotted line) for $N_{c}^{eff}=1.11$ and max(min) CKM matrix elements. Dashed 
line (dot-dashed line) for $N_{c}^{eff}=1.80$ and max(min) CKM matrix elements.}
\label{fig7}
\end{figure}

%
%
A previous analysis~\cite{refa3} has been conducted showing the dependence on the CKM matrix elements and form factors
of the direct $CP$ violating asymmetry. Here, we update our  investigation by taking into account
 the latest values
of the Wolfenstein CKM parameters, $\rho$ and $\eta$, and also by analysing more $B$ decays.
In the following numerical calculations, we apply  the formalism detailed previously and  
investigate   $\bar{B}^{0} \rightarrow \pi^{+} 
\pi^{-} \pi^{0}$ more precisely. We find that for a fixed $N_{c}^{eff}$  there is a maximum
value, $a_{max}$, for the $CP$ violating parameter, $a_{CP}$, when the invariant mass of the $\pi^{+} \pi^{-}$ pair is in
the vicinity of the $\omega$ resonance. In Figs.~\ref{fig6} and~\ref{fig7}, $CP$ violating asymmetries for 
$\Bar{B}^{0} \rightarrow \pi^{+} \pi^{-} \pi^{0}$, for $k^{2}/m_{b}^{2}=0.3$ with  $N_{c}^{eff}=1.09(1.68)$,  
and $k^{2}/m_{b}^{2}=0.5$ with 
$N_{c}^{eff}=1.11(1.80)$, are plotted, respectively, and for limiting values of CKM matrix elements. Graphic 
 results are 
shown only for the model $(1)$, as an example.  We have investigated  five models, with  five different form factors 
in order to test  the model dependence of $a_{CP}$.

Concerning the maximum $CP$ violating asymmetry for $\Bar{B}^{0} \rightarrow \pi^{+} \pi^{-} \pi^{0}$, $a_{max}$, it
varies from
$-51\%(-38\%)$  to  $-84\%(-69\%)$ in the allowed range of $\rho, \eta$ for $k^{2}/m_{b}^{2}=0.3(0.5)$. From
 the numerical
results listed in Table~\ref{tab7}, for $N_{cmin}^{eff}=1.09(1.11)$ and $N_{cmax}^{eff}=1.68(1.80)$, we can see
 that the five
models fall into two classes:  models $(1,3)$ and $(5)$ and models $(2)$ and $(4)$. For models  $(1,3)$ 
and $(5)$, and for 
$N_{cmin}^{eff}=1.09(1.11)$, the maximum asymmetry, $a_{max}$, is around $-54\%(-40\%)$ for the set 
$(\rho_{max},\eta_{max})$ and around $-69\%(-53.6\%)$ for the set $(\rho_{min},\eta_{min})$, leading to the  ratio
between them  being around $1.28(1.34)$. In each of these models and  for $N_{cmax}^{eff}=1.68(1.80)$, the maximum
 value of
 the asymmetry, $a_{max}$, varies from 
$-62.6\%(-48.6\%)$ for the set $(\rho_{max},\eta_{max})$ to around $-77.3\%(-64.6\%)$ for the set 
$(\rho_{min},\eta_{min})$. In that case, the ratio is equal to $1.23(1.32)$. If we consider models $(2)$ and $(4)$,
the maximum  asymmetry, $a_{max}$, where $N_{cmin}^{eff}=1.09(1.11)$, is around $-63.5\%(-48\%)$ for the set 
$(\rho_{max},\eta_{max})$ and around $-78.5\%(-62\%)$ for the set $(\rho_{min},\eta_{min})$. This yields  a ratio
 $1.24(1.29)$. When $N_{cmax}^{eff}=1.68(1.80)$, one has a maximum  asymmetry  around $-71\%(-56.5\%)$ 
for the set $(\rho_{max},\eta_{max})$ and around $-84\%(-69\%)$ for the set $(\rho_{min},\eta_{min})$, leading to a 
 ratio around $1.18(1.22)$.

From all these results, many comments can be enumerated. Although the maximum asymmetry, $a_{max}$, still varies
over  some range in the  $\bar{B}^{0} \rightarrow \pi^{+} \pi^{-} \pi^{0}$ decay, we  stress  that by using
 more accurate CKM element values than before, a more precise $CP$ violating asymmetry is obtained. The reason 
is primarily  the matrix elements $V_{td}$ and $V_{ub}$ which are involved in the $b \rightarrow d$ transition
 through the ratio  of  $p_{\omega}$  to $t_{\rho}$. In our previous $CP$ violation study~\cite{refa3} for the 
 process   $B^{-} \rightarrow \pi^{+}  \pi^{-} \pi^{-}$, we found
that the ratio between the maximum and minimum asymmetry, related to the minimum and maximum set of $(\rho, \eta)$,
 was around $1.6$. By comparison, in the present work, this ratio is reduced to  $1.3$. The difference is related 
 to the improvement in the measurement of the CKM matrix elements, and  shows the strong effect of the 
CKM parameters, $\rho$ and $\eta$, on limiting asymmetry values.

With regard to the  CKM matrix elements, it appears
that if we take their upper limit, we obtain a smaller asymmetry, $a_{CP}$, and vice-versa. As we found  before, 
there is still  a strong dependence of the $CP$ violating asymmetry on the form factors.
The difference between the two classes of models, $(1,3,5)$ and $(2,4)$, comes mainly from the magnitudes of the 
form factors. In fact, the form factor  $F_{1}(k^{2})$, which describes the transition
$B \rightarrow \pi$, is mainly responsible for  this dependence. In both classes, we find a stronger dependence of the 
$CP$ violating asymmetry on the CKM matrix elements than that on the form factors or the effective parameter 
$N_{c}^{eff}$. The difference
observed in our results between $k^{2}/m_{b}^{2}=0.3$ and  $k^{2}/m_{b}^{2}=0.5$ arises from the $k^{2}$ dependence
of the Wilson coefficients in the weak effective Hamiltonian. Finally, since $N_{c}^{eff}$ (treated as a free 
parameter) is related  to
 hadronization effects through the  factorization approach, it is not possible to determine its value accurately 
(since  non-factorizable effects are  not well known). That is   why
 the asymmetry also varies in some range of   $N_{c}^{eff}$. It is obvious
 that a more accurate  value for  $N_{c}^{eff}$  (which requires a more accurate approach with non-factorizable 
effects being taken into account), and  hadronic decay form factors (which requires better understanding for 
 pionic  structure  and  the $B \rightarrow \pi$ transition) are needed in order to determine  
 the CKM matrix elements.

\begin{figure}
\centering\includegraphics[height=7cm,clip=true]{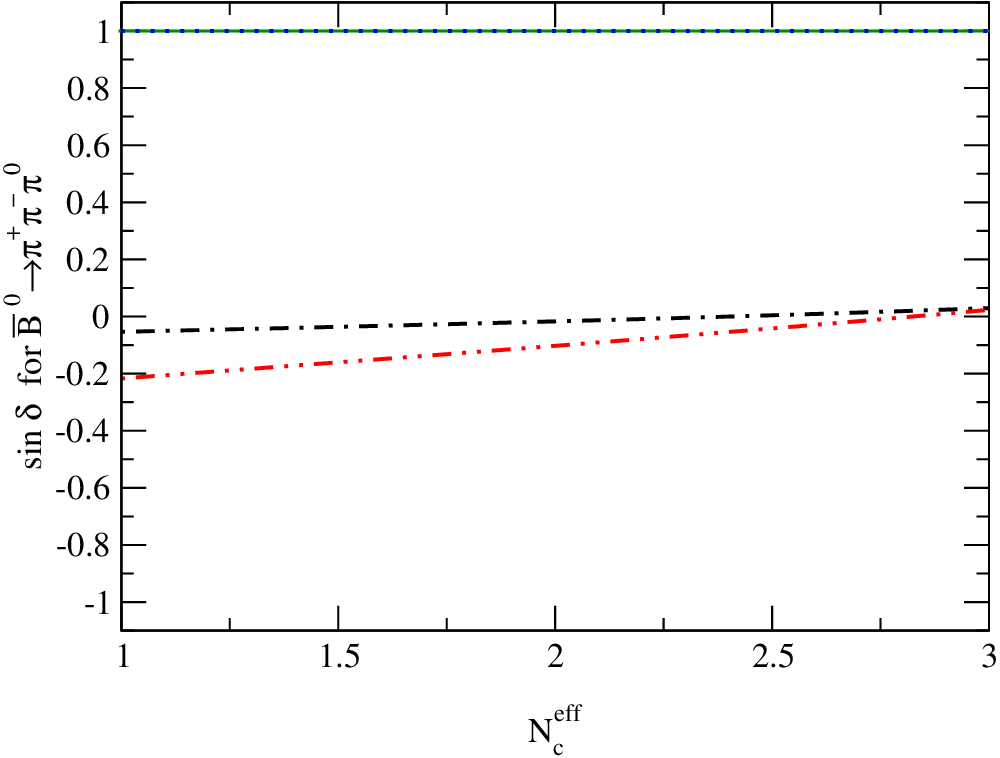}
\caption{$\sin\delta$ as a function of $N_{c}^{eff}$, for $\bar{B}^{0} \rightarrow
 \pi^{+} \pi^{-} \pi^{0}$, for $k^{2}/m_{B}^{2}=0.3(0.5)$ and for  model $(1)$.   The solid (dotted) line at
 $\sin \delta=+1$ corresponds the case    $\tilde{\Pi}_{\rho \omega}=(-3500;-300)$, where     
 $\rho - \omega$ mixing is included. The  dot-dashed (dot-dot-dashed) line corresponds to 
 $\tilde{\Pi}_{\rho \omega}=(0;0)$, where   $\rho - \omega$ mixing is not included. }
\label{fig10}
\end{figure}

\begin{figure}
\centering\includegraphics[height=7cm,clip=true]{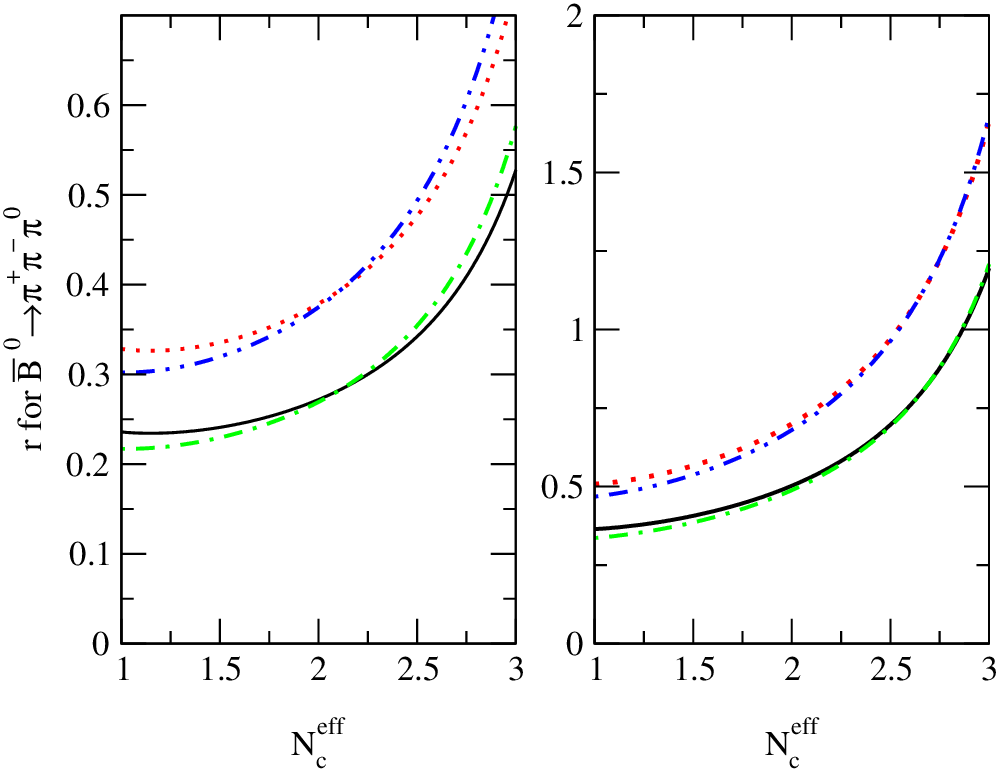}
\caption{The ratio of penguin to tree amplitudes, $r$, as a function of $N_{c}^{eff}$, for $\bar{B}^{0} \rightarrow 
\pi^{+} \pi^{-} \pi^{0}$, for $k^{2}/m_{B}^{2}=0.3(0.5)$, 
for limiting values of the CKM matrix elements $(\rho,\eta)$ max(min),  for  $\tilde{\Pi}_{\rho
 \omega}=(-3500;-300)(0,0)$, (i.e. with(without) $\rho - \omega$  mixing) and for   model $(1)$. Figure 4a
 (left): for $\tilde{\Pi}_{\rho \omega}=(0;0)$,  solid line (dotted line) for $k^{2}/m_{B}^{2}=0.3$ and
 $(\rho,\eta)$ max(min). Dot-dashed line (dot-dot-dashed line) for  $k^{2}/m_{B}^{2}=0.5$ and $(\rho,\eta)$
 max(min). Figure 4b (right): same caption but for  $\tilde{\Pi}_{\rho \omega}=(-3500;-300)$. }
\label{fig12}
\end{figure}

In spite of all the uncertainties mentioned above, we stress that the $\rho-\omega$ mixing mechanism in 
the $B \rightarrow \rho \pi$ decay can be used  to remove  ambiguity concerning the sign of $\sin \delta$. As the
internal top quark dominates the $b \rightarrow d$ transition, the weak phase in the  asymmetry is proportional
to $\sin \alpha \; (=\sin \phi)$, where $\alpha={\rm arg}\left[ - \frac{V_{td}V_{tb}^{\star}}{V_{ud}
V_{ub}^{\star}}\right]$.  Hence  knowing the sign of $\sin \delta$ enables us to determine  that  of $\sin \alpha$ 
from a measurement of the asymmetry, $a_{CP}$. In Fig.~\ref{fig10}  we show  $\sin \delta$
as a function of $N_{c}^{eff}$  for $\bar{B}^{0} \rightarrow \pi^{+} \pi^{-} \pi^{0}$  when we have maximum $CP$
 violation. Then, in our determined range of $N_{c}^{eff}$, ($1.09(1.11)
< N_{c}^{eff}<1.68(1.80)$), one finds
that its sign is always positive for all the models studied and for all the form factors. Therefore, by measuring the
$CP$ violating asymmetry in  $\bar{B}^{0} \rightarrow \pi^{+} \pi^{-} \pi^{0}$, we can remove
 the  mod($\pi$) ambiguity which appears in the determination for $\alpha$ from the 
usual indirect measurements  which yield $\sin 2 \alpha$. 
In Fig.~\ref{fig12}, the  ratio of the  penguin and  tree amplitudes, as a function
of $N_{c}^{eff}$, is plotted for  limiting values of the CKM matrix elements, $\rho, \eta$, for the process
$\Bar{B}^{0} \rightarrow \pi^{+} \pi^{-} \pi^{0}$. Even though one gets a larger value of $\sin \delta$
around $N_{c}^{eff}=1$, for $\bar{B}^{0} \rightarrow \pi^{+} \pi^{-} \pi^{0}$, without $\rho-\omega$ mixing, one still
has a small value for $r$ around this value of $N_{c}^{eff}$.  In that case, the $CP$ violating
asymmetry, $a_{CP}$, remains very small without $\rho-\omega$ mixing.
%
%
\section{Branching ratios for  $\boldsymbol{B^{\pm,0} \rightarrow \rho^{0}\pi^{\pm,0}}$}\label{part5}
%
%
\subsection{Formalism}\label{part5.1}
%
The direct   $B  \rightarrow \rho^{0} \pi$ transition 
is the  main contribution to the decay rate. In our case, to be consistent, we should also take into account the 
 $\rho - \omega$ mixing contribution   to the branching ratio, since we are working to the first order of  isospin
 violation. The derivation  is straightforward and we obtain the following form for the branching ratio for 
$B \rightarrow \rho^{0} \pi$:
\begin{multline}\label{eq1.53}
BR(B  \rightarrow \rho^{0} \pi )=\frac{G_{F}^{2}|\vec{p}_{\rho}|^{3}}{\alpha_k \pi \Gamma_{B}}\Bigg|
\bigg[V_{d}^{T}A^{T}_{\rho^{0}}(a_{1},a_{2})-V_{d}^{P}A^{P}_{\rho^{0}}(a_{3}, \cdots, a_{10})\bigg]    \\
+\bigg[V_{d}^{T}A^{T}_{\omega}(a_{1},a_{2})-V_{d}^{P}A^{P}_{\omega}(a_{3}, \cdots, a_{10})\bigg]\frac{
\tilde{\Pi}_{\rho \omega}}{(s_{\rho}-m_{\omega}^{2})+im_{\omega}\Gamma_{\omega}}\Bigg|^{2}\ .
\end{multline}
In Eq.~(\ref{eq1.53}) $G_{F}$ is the Fermi constant, $\Gamma_{B}$ is  the $B$ total decay width, and $\alpha_k$ is 
an integer related
to the given decay, $A^{T}_{V}$ and $A^{P}_{V}$ are the tree and penguin amplitudes, and   $V_{d}^{T},V_{d}^{P}$ 
 represent the CKM matrix elements involved in  the tree and penguin diagrams, respectively: 
\begin{equation}
V_{d}^{T}  =|V_{ub}V_{ud}^{\star}|\ ,  \;\;\;\; {\rm and} \;\;\; 
V_{d}^{P}  =|V_{tb}V_{td}^{\star}|\ .
\end{equation}
\noindent The effective parameters, $a_{i}$, which are involved in the decay amplitude,  are the following
combinations of effective Wilson coefficients:
\begin{equation}\label{eq1.44b}
a_{2j}=C_{2j}^{\prime}+\frac{1}{N_{c}^{eff}}C_{2j-1}^{\prime},   \;\;\; a_{2j-1}=C_{2j-1}^{\prime}+\frac{1}
{N_{c}^{eff}}C_{2j}^{\prime}, \;\;  {\rm for} \;\; j=1, \cdots, 5\ .
\end{equation}
%

%
%
%
\subsection{Calculational details}\label{part5.2}
%
%
In this section, we give full details of the  theoretical decay amplitudes for decays involving the 
  $b$ to $d$ transition. Two of these  decays  involve $\rho-\omega$ mixing. They are
$B^{-} \rightarrow \rho^{0} \pi^{-}$ and $\bar{B}^{0} \rightarrow \rho^{0} \pi^{0}$. The other two decays are 
$\bar{B}^{0} \rightarrow \rho^{-} \pi^{+}$ and $B^{-} \rightarrow \rho^{-} \pi^{0}$. We list  in the following, the
tree and penguin amplitudes which appear in the given transitions.
\newline
\noindent For the decay  $B^{-} \rightarrow \rho^{0} \pi^{-}$ ($\alpha_k = 32$ in Eq.~(\ref{eq1.53})),
\begin{multline}\label{eq1.55}
\sqrt{2}A^{T}_{\rho}(a_{1},a_{2})  = a_{1}f_{\rho}F_{1}(m_{\rho}^{2})+a_{2}f_{\pi}A_{0}(m_{\pi}^{2})\ , 
\hspace{16em}
\end{multline}
\begin{multline}\label{eq1.56}
\sqrt{2}A^{P}_{\rho}(a_{3},  \cdots,   a_{10})  =
f_{\rho}F_{1}(m_{\rho}^{2}) \biggl\{ -a_{4} +\frac{3}{2}(a_{7}+a_{9}) + \frac{1}{2}a_{10} \biggr\} \\
+f_{\pi}A_{0}  (m_{\pi}^{2})\Biggl\{ a_{4} -2(a_{6}+a_{8}) \biggl[ \frac{m_{\pi}^{2} }
{(m_{u}+m_{d})(m_{b}+m_{u})}\biggr] + a_{10} \Biggr\}\ ;
\end{multline}
\noindent for the decay  $B^{-} \rightarrow \omega \pi^{-}$ ($\alpha_k = 32$ in Eq.~(\ref{eq1.53})),
\begin{multline}\label{eq1.57}
\sqrt{2}A^{T}_{\omega}(a_{1},a_{2})  = a_{1}f_{\rho}F_{1}(m_{\rho}^{2})+a_{2}f_{\pi}A_{0}(m_{\pi}^{2})\ ,
 \hspace{16em}
\end{multline}
\begin{multline}\label{eq1.58}
\sqrt{2}A^{P}_{\omega}(a_{3},  \cdots,   a_{10})  = f_{\rho} F_{1}(m_{\rho}^{2})
 \biggl\{ 2(a_{3}+a_{5})+\frac{1}{2}(a_{7}
+a_{9}) + (a_{4} -\frac{1}{2} a_{10})  \biggr\} \\
 + f_{\pi}A_{0}(m_{\pi}^{2})\Biggl\{ - 2 (a_{6}+a_{8}) \biggl[ \frac{m_{\pi}^{2}}{(m_{u}+m_{d})(m_{b}+m_{u})}\biggr]
 + a_{4} + a_{10} \Biggr\}\ ;  
\end{multline}
\noindent for the decay  $\bar{B}^{0} \rightarrow \rho^{0} \pi^{0}$ ($\alpha_k = 64 $ in Eq.~(\ref{eq1.53})),
\begin{multline}\label{eq1.59}
 2 A^{T}_{\rho}(a_{1},a_{2})  = a_{1} f_{\rho}F_{1}(m_{\rho}^{2})+ a_{1} f_{\pi}A_{0}(m_{\pi}^{2})\ , 
 \hspace{17em} 
\end{multline}
\begin{multline}\label{eq1.60}
2 A^{P}_{\rho}(a_{3},  \cdots,   a_{10})  = 
f_{\rho}F_{1}(m_{\rho}^{2}) \biggl\{ -a_{4} + \frac{1}{2}(3 a_{7}+3 a_{9} +a_{10}) \biggr\}+ \\
f_{\pi} A_{0}(m_{\pi}^{2}) \Biggl\{ -a_{4} + (2 a_{6} - a_{8}) \biggl[ \frac{m_{\pi}^{2}}
{2 m_{d}(m_{b}+ m_{d})}\biggr] + \frac{1}{2}(-3 a_{7}+3 a_{9} + a_{10}) \Biggr\}\ ;
\end{multline}
\noindent for the decay  $\bar{B}^{0} \rightarrow \omega \pi^{0}$ ($\alpha_k = 64$ in Eq.~(\ref{eq1.53})),
\begin{multline}\label{eq1.61}
 2 A^{T}_{\omega}(a_{1},a_{2})  = -a_{1} f_{\rho}F_{1}(m_{\rho}^{2})+ a_{1} f_{\pi}A_{0}(m_{\pi}^{2})\ ,  
\hspace{17em} 
\end{multline}
\begin{multline}\label{eq1.62}
  2 A^{P}_{\omega}(a_{3},  \cdots,   a_{10})  = 
f_{\rho}F_{1}(m_{\rho}^{2}) \biggl\{- 2 (a_{3}+ a_{5}) -a_{4} - \frac{1}{2}(a_{7}+a_{9}-a_{10})\biggr\} \\
+f_{\pi} A_{0}(m_{\pi}^{2}) \Biggl\{- a_{4} + (2 a_{6} - a_{8}) \biggl[ \frac{m_{\pi}^{2}}
{2 m_{d}(m_{b}+ m_{d})}\biggr] + \frac{1}{2}(-3 a_{7}+3 a_{9}+ a_{10}) \Biggr\}\ ;  
\end{multline}
\noindent for the decay $\bar{B}^{0} \rightarrow \rho^{-} \pi^{+}$ ($\alpha_k = 16$ in Eq.~(\ref{eq1.53})),
\begin{multline}\label{eq1.63}
A^{T}_{\rho}(a_{1},a_{2})  = a_{2}f_{\rho}F_{1}(m_{\rho}^{2})\ , 
\hspace{23em}
\end{multline}
\begin{multline}\label{eq1.64}
A^{P}_{\rho}(a_{3},\cdots, a_{10})  = (a_{4}+a_{10})f_{\rho}F_{1}(m_{\rho}^{2})\ ; 
\hspace{18em}
\end{multline}
\noindent for the decay $B^{-} \rightarrow \rho^{-} \pi^{0}$ ($\alpha_k = 32$ in Eq.~(\ref{eq1.53})),
\begin{multline}\label{eq1.65}
{\sqrt 2} A^{T}_{\rho}(a_{1},a_{2})  = a_{2}f_{\rho}F_{1}(m_{\rho}^{2}) + a_{1} f_{\pi} A_{0}(m_{\pi}^{2})\ , 
\hspace{16em}
\end{multline}
\begin{multline}\label{eq1.66}
{\sqrt 2} A^{P}_{\rho}(a_{3},\cdots, a_{10})  = f_{\rho}F_{1}(m_{\rho}^{2}) (a_{4} +a_{10}) + \\
f_{\pi} A_{0}(m_{\pi}^{2}) \Biggl\{ -a_{4} - \frac{1}{2}(3 a_{7}-3 a_{9}-a_{10}) +(2 a_{6} - a_{8}) 
\biggl[ \frac{m_{\pi}^{2}} { 2 m_{d}(m_{b}+ m_{d})} \biggr] \Biggr\} \ . 
\end{multline}
Moreover, we can calculate the ratio between  two branching ratios, namely 
$BR(B^{0} \rightarrow \rho^{\pm} \pi^{\mp})$ and $BR(B^{\pm} \rightarrow \rho^{0} \pi^{\pm})$,  in which
 the uncertainty caused by many systematic errors is removed.  We define the ratio, $R_{\pi}$,  as: 
\begin{equation}\label{eq1.67}
R_{\pi}= \frac{BR(B^{0} \rightarrow \rho^{\pm} \pi^{\mp})}{BR(B^{\pm} \rightarrow \rho^{0} \pi^{\pm})}\ .
\end{equation}
%
%

%
\subsection{Numerical results}\label{part5.3}
%

The numerical values for the  CKM matrix elements  $V_{d}^{T,P}$, 
$\rho-\omega$ mixing amplitude  $\tilde{\Pi}_{\rho \omega}$, and   particle masses  $m_{V,P}$, which appear  in 
Eq.~(\ref{eq1.53}),  have been 
  reported in Sections~\ref{part2.1} and~\ref{part3}. The Fermi constant
is taken to be  $G_{F}= 1.166391 \times 10^{-5}  {\rm GeV}^{-2}$~\cite{ref16}, and for the total decay width $B$ meson, 
$\Gamma_{B}(= 1/\tau_{B})$,
we use  the  world average $B$ life-time values (combined results from ALEPH, CDF, DELPHI, L3, OPAL and  
SLD)~\cite{ref26}:
\begin{align}\label{eq1.69}
\tau_{B^{0}} & = 1.546 \pm 0.021 \; {\rm ps}\ , \nonumber \\
\tau_{B^{+}} & = 1.647 \pm 0.021 \; {\rm ps}\ .
\end{align}

To compare theoretical results with experimental data,  as well as to  determine constraints on the effective number
of colours, $N_{c}^{eff}$, the form factors and the  CKM matrix parameters, we shall use the experimental
 branching ratios
collected by  CLEO~\cite{ref29}, BELLE~\cite{ref30,ref31,ref32} and BABAR~\cite{ref33,ref34} factories. All the
experimental values are summarized  in Table~\ref{tab6}.

\begin{figure}
\centering\includegraphics[height=7cm,clip=true]{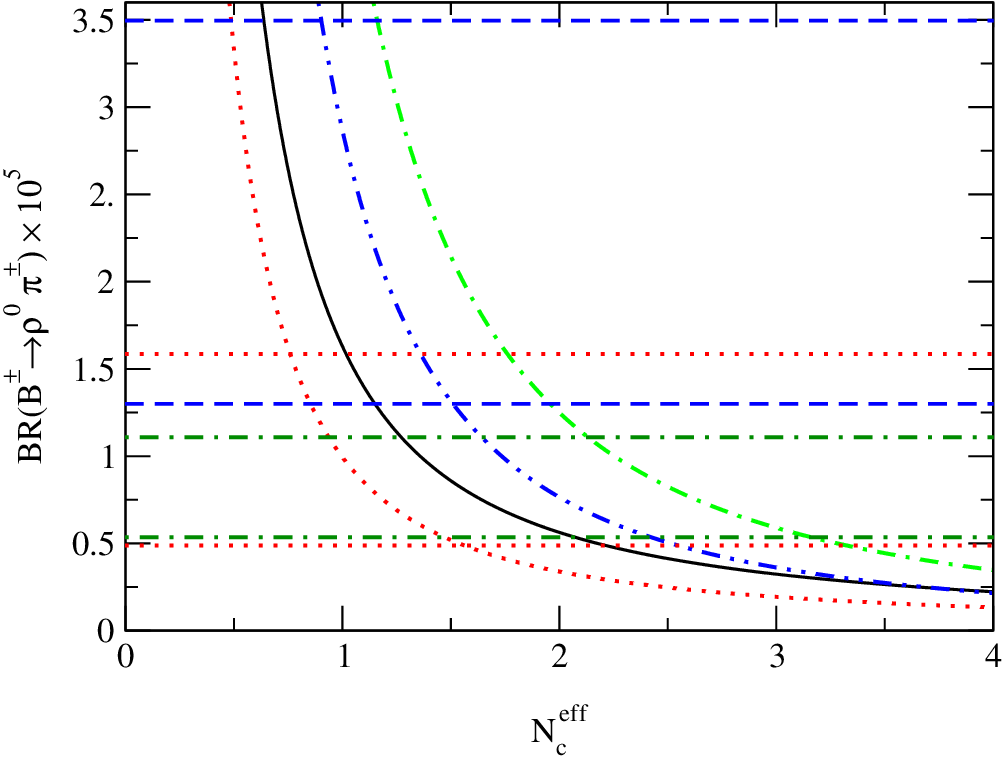}
\caption{Branching ratio for $B^{\pm} \rightarrow \rho^{0}  \pi^{\pm}$ for models  $1(2)$, $k^{2}/m_{B}^{2}=0.3$ 
and limiting values of the  CKM matrix elements. Solid line (dotted line) for  model $(1)$ and max(min) CKM matrix 
elements. Dot-dashed line (dot-dot-dashed line) for   model $(2)$ and max(min) CKM matrix elements. 
Horizontal 
dotted  lines: CLEO data; horizontal dashed lines: BABAR data; horizontal dot-dashed lines: BELLE data.}
\label{fig14}
\end{figure}

\begin{figure}
\centering\includegraphics[height=7cm,clip=true]{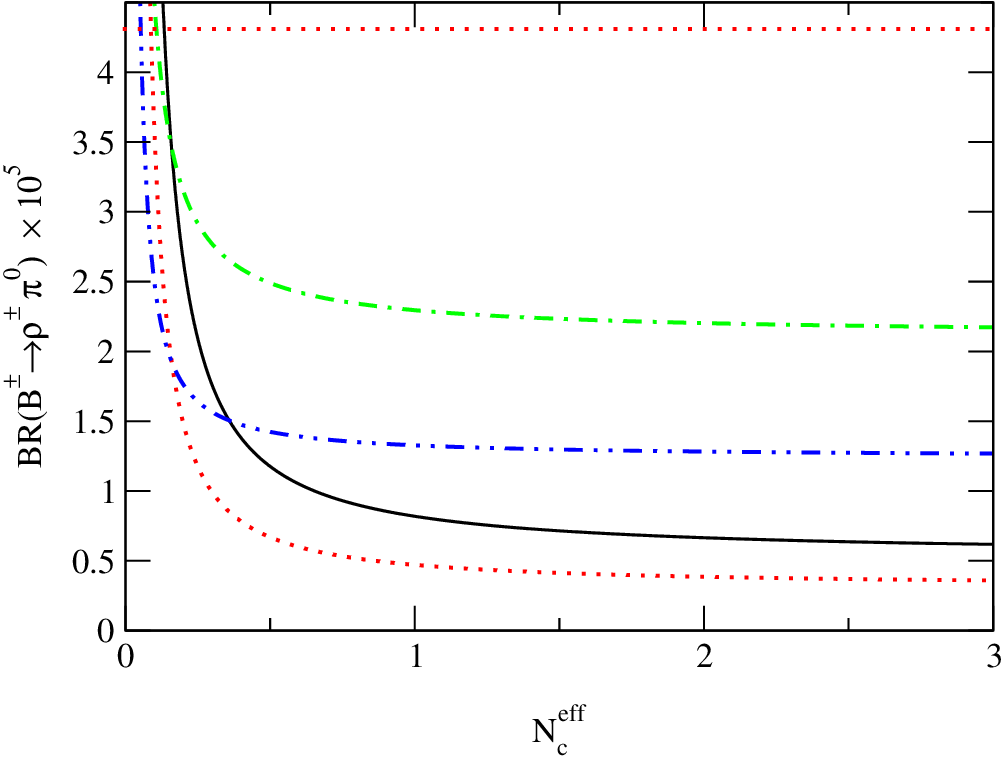}
\caption{Branching ratio for $B^{\pm} \rightarrow \rho^{\pm}\pi^{0}$ for models  $1(2)$, $k^{2}/m_{B}^{2}=0.3$
 and limiting values of the  CKM matrix elements. Solid line (dotted line) for  model $(1)$ and max(min) CKM 
matrix elements. Dot-dashed line (dot-dot-dashed line) for  model $(2)$ and max(min) CKM matrix elements. 
Same notation for experimental data as in  Fig.~\ref{fig14}.}
\label{fig15}
\end{figure}

\begin{figure}
\centering\includegraphics[height=7cm,clip=true]{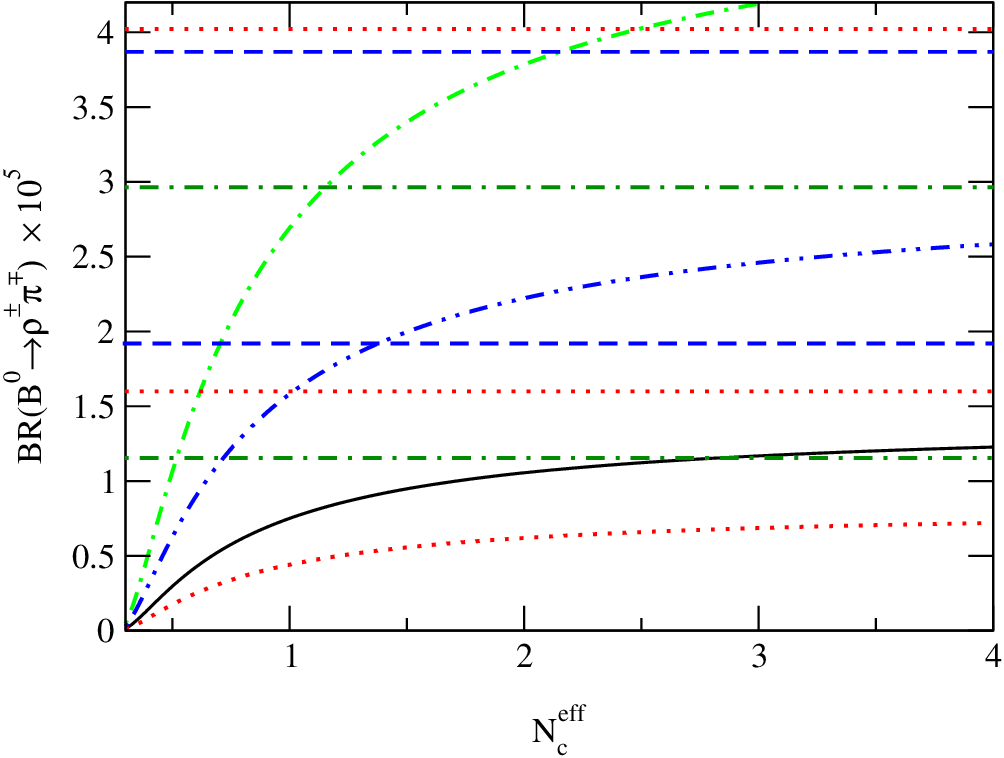}
\caption{Branching ratio for $B^{0} \rightarrow \rho^{\pm}\pi^{\mp}$ for models  $1(2)$, $k^{2}/m_{B}^{2}=0.3$
 and limiting values of the  CKM matrix elements. Solid line (dotted line) for  model $(1)$ and max(min) CKM 
matrix elements. Dot-dashed line (dot-dot-dashed line) for  model $(2)$ and max(min) CKM matrix elements. 
Same notation for experimental data as in Fig.~\ref{fig14}.}
\label{fig16}
\end{figure}

\begin{figure}
\centering\includegraphics[height=7cm,clip=true]{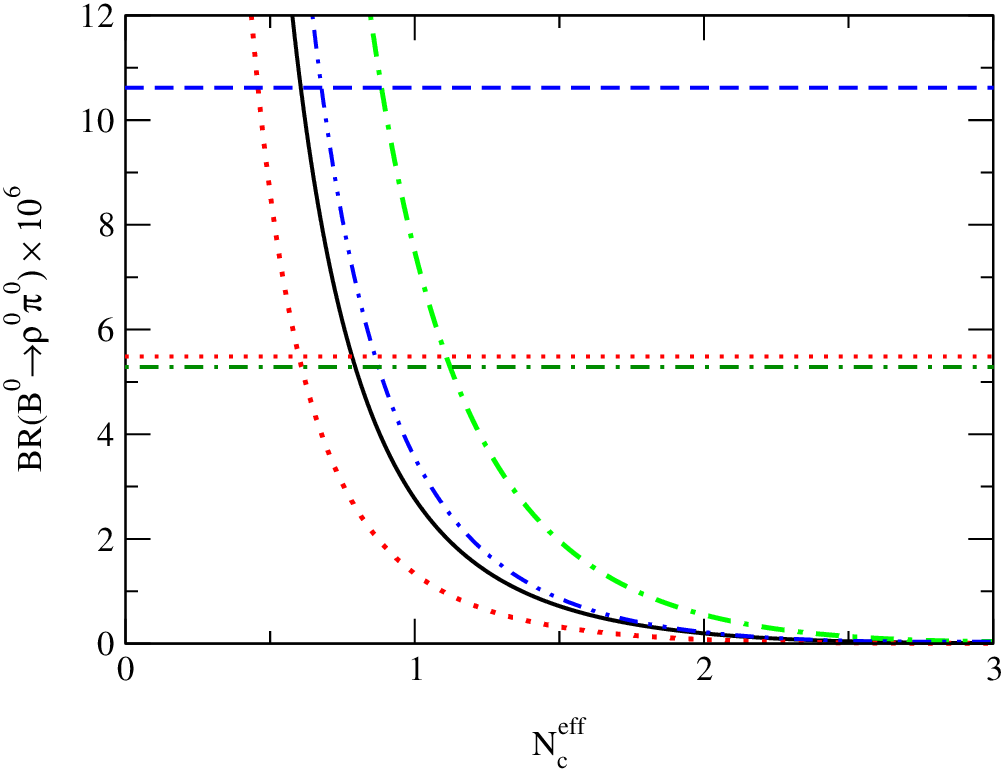}
\caption{Branching ratio for $B^{0} \rightarrow \rho^{0}\pi^{0}$ for models  $1(2)$, $k^{2}/m_{B}^{2}=0.3$
 and limiting values of the  CKM matrix elements. Solid line (dotted line) for  model $(1)$ and max(min) CKM 
matrix elements. Dot-dashed line (dot-dot-dashed line) for  model $(2)$ and max(min) CKM matrix elements. 
Same notation for experimental data as in Fig.~\ref{fig14}.}
\label{fig17}
\end{figure}

In order to determine the range of $N_{c}^{eff}$, which is allowed by experimental data, we have calculated
 the branching
ratios for $B^{\pm} \rightarrow \rho^{0} \pi^{\pm}$, $B^{\pm} \rightarrow \rho^{\pm} \pi^{0}$, 
 $B^{0} \rightarrow \rho^{\pm} \pi^{\mp}$, and $B^{0} \rightarrow \rho^{0} \pi^{0}$. All the results are shown
in Figs.~\ref{fig14},~\ref{fig15},~\ref{fig16} and~\ref{fig17} for the corresponding branching ratios listed
above. Results are plotted for models $(1)$ and $(2)$, since they involve  different form factor values and 
thus  show their dependence on form factors. As experimental data, we shall use three sets of data from the
 CLEO, BABAR and BELLE Collaborations, respectively. Since experimental branching ratios from CLEO are the most
 accurate, we shall
use them to extract the range of $N_{c}^{eff}$. The other two, the BABAR and BELLE data, will give us an idea of
 the magnitude
of the experimental uncertainties. It is clear that numerical results are very sensitive to uncertainties coming from
the experimental data. Thus, the determination of the allowed range of $N_{c}^{eff}$ will be done by using  all the 
branching ratio results.

\begin{figure}
\centering\includegraphics[height=7cm,clip=true]{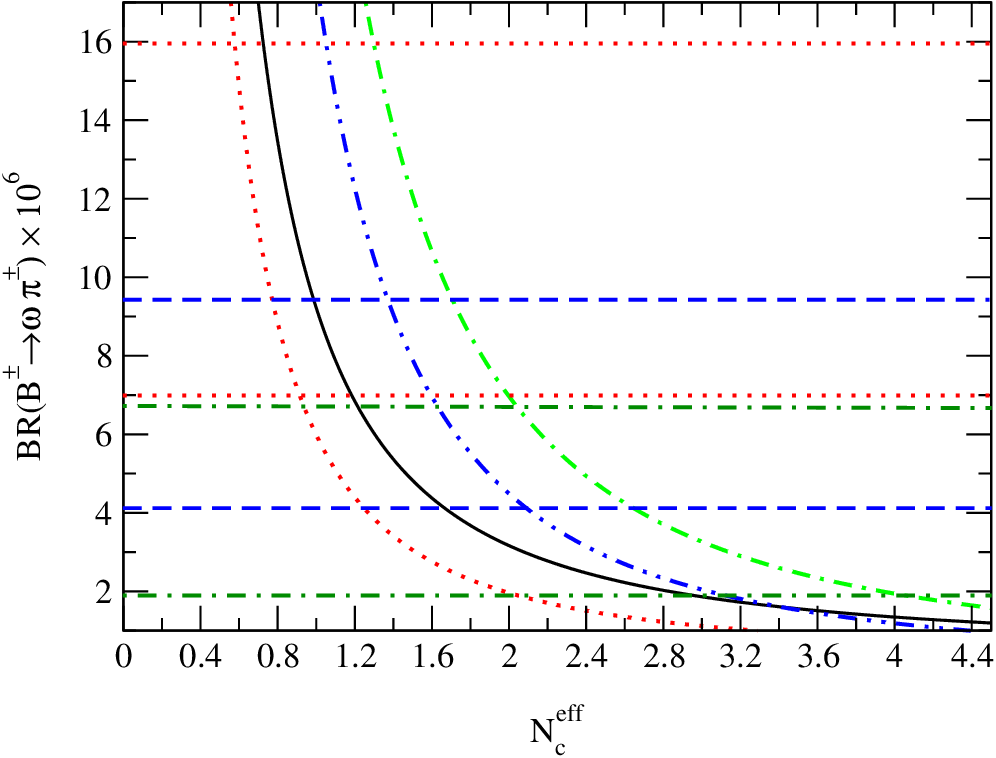}
\caption{Branching ratio for $B^{\pm} \rightarrow \omega \pi^{\pm}$ for models  $1(2)$, $k^{2}/m_{B}^{2}=0.3$
 and limiting values of the  CKM matrix elements. Solid line (dotted line) for  model $(1)$ and max(min) CKM 
matrix elements. Dot-dashed line (dot-dot-dashed line) for  model $(2)$ and max(min) CKM matrix elements. 
Same notation for experimental data as in  Fig.~\ref{fig14}.}
\label{fig18}
\end{figure}

Let us start with the decay processes  $B^{-} \rightarrow \rho^{0} \pi^{-}$ and $B^{-} \rightarrow 
\rho^{-} \pi^{0}$. In both cases, there is a large range of acceptable values 
 for  $N_{c}^{eff}$ and the CKM matrix elements 
over which the theoretical results are consistent with experimental data from CLEO, BABAR and BELLE. For $B^{-} 
\rightarrow \rho^{-} \pi^{0}$, the lack of data does not allow  us to determine the range. However, 
experiment and theory are consistent in  both  cases. For $B^{-} \rightarrow \rho^{0} \pi^{-}$,  the 
models show considerable variation
 even though they are all consistent with the experimental data. Numerical results for models $(1,3)$ and
 $(5)$ are close, so  are 
those for models $(2)$ and $(4)$. We emphasise that the
 effect of $\rho-\omega$ mixing on the branching ratio $B^{\pm} \rightarrow \rho^{0} \pi^{\pm}$ can be as large as
 $30\%$.
As regards  $B^{0} \rightarrow \rho^{-} \pi^{+}$ and $\bar{B}^{0} \rightarrow \rho^{0} \pi^{0}$, the results 
and conclusions
are different from those for $B^{-} \rightarrow \rho^{0} \pi^{-}$. If we look at the branching ratio
 for $B^{0} \rightarrow 
\rho^{\pm} \pi^{\mp}$, only models $(2)$ and $(4)$ are consistent with experimental data over a large range of 
$N_{c}^{eff}$, whereas models $(1,3)$ and $(5)$ are not. The strong sensitivity to the results in that case comes from
the fact that the decay branching ratios for $B^{0} \rightarrow  \rho^{\pm} \pi^{\mp}$ depend on form factors
more sensitively, because  in this case only one form
factor, $F_{1}(k^{2})$, is involved. In all the other cases,
the amplitudes depend on  both $F_{1}(k^{2})$ and $A_{0}(k^{2})$.
Therefore these branching ratios are less sensitive to the magnitude of the form factors. Finally, for the 
branching ratio $BR(B^{\pm} \rightarrow \omega \pi^{\pm})$ plotted in 
Fig.~\ref{fig18}, all models give theoretical results  consistent with  experimental data. Once again, the 
difference observed between models $(1)$ and $(2)$ mainly comes from the form factor $F_{1}(k^{2})$ (i.e. from the 
pion wave function used). Our analysis shows that models $(1,3)$ and $(5)$ cannot give  results consistent 
with all experiments and have to be excluded.

\begin{figure}
\centering\includegraphics[height=7cm,clip=true]{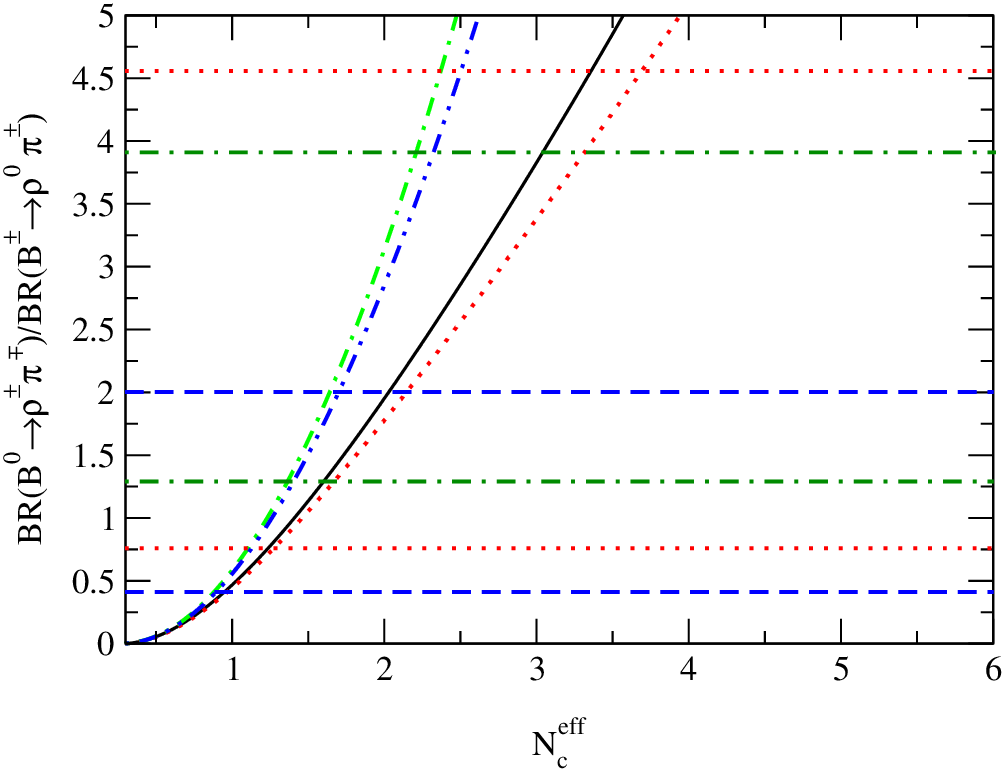}
\caption{The ratio of two $\rho \pi$ branching ratios versus $N_{c}^{eff}$  for models $1(2)$ 
and for limiting values of the CKM matrix elements: solid  line (dotted line) for  model $(1)$ with max(min) CKM
 matrix elements.  Dot-dashed line (dot-dot-dashed line) for   model $(2)$ with max(min) CKM matrix elements. 
Same notation for experimental data  as in Fig.~\ref{fig14}.}
\label{fig19}
\end{figure}

To remove systematic uncertainties coming from experimental results, one can calculate the ratio between two 
branching ratios for $B$ decays.
In the present case (with the data available), the ratio, $R_{\pi}$, is between $BR(B^{\pm} \rightarrow 
\rho^{0} \pi^{\pm})$ 
and  $BR(B^{0} \rightarrow \rho^{\pm} \pi^{\mp})$. Results are shown in Fig.~\ref{fig19}. We observe that the ratios
differ totally from each other for  models $(1,3)$ and $(5)$ and models $(2)$ and $(4)$. Since models $(1,3)$ and $(5)$
 have already
been excluded, we will use models $(2)$ and $(4)$ for the determination of the range for $N_{c}^{eff}$. If we 
just include tree
contributions in the decay amplitudes,  $R_{\pi}$ becomes independent of the CKM matrix elements. Penguin contributions
lead to  a relatively weak dependence of $R_{\pi}$ 
on the CKM matrix elements. By comparing numerical results and 
experimental data, we are now able to extract a  range for   $N_{c}^{eff}$ which is  consistent with
all the  results. 
To determine the best range of $N_{c}^{eff}$, we 
select the values of  $N_{c}^{eff}$ which are allowed by all constraints for each model.
Finally, after excluding models $(1,3)$ and $(5)$ for the  obvious reasons
mentioned before, we can now fix
the upper and the lower limit of the range of $N_{c}^{eff}$ (Table~\ref{tab15}). 
We find that $N_{c}^{eff}$ should be in the range
$1.09(1.11)< N_{c}^{eff}<1.68(1.80)$ for $k^{2}/m_{b}^{2}=0.3(0.5)$. Comparing with our previous study, the current
 range of  $N_{c}^{eff}$ is consistent but  smaller  than the previous one.
%
%
\section{Summary and discussion}\label{part6}
%
%
The first aim of the present work was to compare theoretical branching ratios for $B^{\pm} \rightarrow \rho^{0} 
\pi^{\pm}$, $B^{\pm} \rightarrow 
\rho^{\pm} \pi^{0}$,  $B^{0} \rightarrow \rho^{\pm} \pi^{\mp}$ and  $B^{0} \rightarrow \rho^{0} \pi^{0}$ with
 experimental data from the CLEO, BABAR and BELLE Collaborations. The second 
was to apply recent values of the CKM matrix elements, e.g. $A, \lambda, \eta$ and $\rho$, to  study 
direct $CP$ violation for $B$ decay such as $\Bar{B}^{0} \rightarrow \rho^{0}(\omega) \pi^{0} \rightarrow \pi^{+}
 \pi^{-} \pi^{0}$, where   the $\rho-\omega$
mixing mechanism must  be included. The advantage of including $\rho-\omega$ mixing is that the strong phase 
difference which is  necessary for  
direct $CP$ violation, is large  and rapidly varying near the $\omega$ resonance. As a result, the $CP$ violating
asymmetry, $a_{CP}$, reaches a maximum, $a_{max}$, when the invariant mass of the $\pi^{+}\pi^{-}$ pair is in the 
vicinity of the $\omega$ resonance and $\sin \delta = +1$ at this point.

In our approach, we started from the weak effective Hamiltonian where short distance and long distance physics
are  separated and treated  by a perturbative approach (Wilson coefficients)  and  a  non-perturbative 
approach (operator
product expansion), respectively. One of the main uncertainties introduced in 
our calculation  comes from  the hadronic matrix elements for both tree and penguin operators. We treated them by
 applying a 
naive factorization approximation, where $N_{c}^{eff}$ is taken as an effective parameter. Although this is  clearly
an approximation, it has been pointed out~\cite{ref36} that it may be quite reliable in energetic weak decays
such as $B \rightarrow \rho \pi$.

We have investigated the direct $CP$ violating asymmetry in the $B$ decay:  $\Bar{B}^{0} \rightarrow \pi^{+} \pi^{-}
 \pi^{0}$. 
We found that the $CP$ violation parameter, $a_{CP}$, is 
very sensitive to the parameters $\rho$ and $\eta$ in 
the CKM matrix, and also to the magnitude of the form factors appearing in the five
 phenomenological models we investigated. We have calculated  the
maximum  asymmetry, $a_{max}$, as a function of the effective parameter, $N_{c}^{eff}$, with the limiting 
values of the CKM  matrix elements. We found that the $CP$ violating asymmetry, $a_{max}$, can vary 
from $-37\%$ to $-84\%$ over all the models $(1,2,3,4,5)$. As we  already suggested in a
previous study~\cite{refa3}, the ratio between the 
asymmetries for   limiting values of the CKM matrix elements is mainly governed by $\eta$. Previously, we found
 a ratio equal to $1.64$ where the CKM values used were the following: $A=0.815, \lambda = 0.2205$, $0.09 
< \rho < 0.254$, and $0.323 < \eta  < 0.442 $. In the present work, we found for the same decay, a ratio
 equal to $1.30$. The more accurate value for  $\eta$ has reduced  uncertainties on both the $CP$ violating
 asymmetry and the ratio, $\Gamma (B^{\pm} \rightarrow \rho^{0} \pi^{\pm})/
\Gamma (B^{0} \rightarrow \rho^{0} \pi^{0})$.

Moreover, 
we stressed that without the
$\rho - \omega$ mixing mechanism, the $CP$ violating asymmetry, $a_{CP}$ (which is  proportional to both $\sin
 \delta$ and $r$), 
is small since in that case  either $\sin \delta$ or $r$ is small. In the allowed range of $N_{c}^{eff}$, we also
found that the sign of $\sin \delta$ is always positive. Therefore, by measuring $a_{CP}$, we can remove the phase 
mod($\pi$) ambiguity  which occurs in the usual method for the determination of the CKM   unitarity angle $\alpha$.

We have calculated branching ratios for $B^{\pm} \rightarrow \rho^{0} \pi^{\pm}$, $B^{\pm} 
\rightarrow \rho^{\pm} \pi^{0}$,  $B^{0} \rightarrow \rho^{\pm} \pi^{\mp}$ and  $B^{0} \rightarrow \rho^{0}
 \pi^{0}$ and compared the results with experimental data coming from the CLEO, BABAR and BELLE Collaborations.
We have shown that for  models $(2)$ and $(4)$ there is a range for $N_{c}^{eff}$, 
 $1.09(1.11)< N_{c}^{eff}<1.68(1.80)$, in which theoretical results are  consistent with experimental data. Models
$(1,3)$ and  $(5)$ are excluded since the  form factor $F_{1}(k^{2})$ in these models cannot produce  results
consistent with experiment. For a deeper investigation into this problem,  some resonant and non-resonant 
contributions~\cite{ref37, ref38} which may carry bigger effects than
expected in the calculation of branching ratios in $\rho \pi$ may have to be considered seriously.

With more   accurate CKM matrix elements values, e.g. $\rho$ and $\eta$, we are   able
to give more  precise  $CP$ violating asymmetries, and   the main uncertainties remaining are from the 
factorization~\cite{ref3a}
approach and the hadronic decay form factors.  In the future one may hope to use QCD factorization
to replace the effective parameter, $N_{c}^{eff}$, and hence to provide a more reliable treatment of non-factorizable
effects. With regard to form factors,  we have shown that some models for the $B \rightarrow \pi$ 
transition are not consistent with the experimental branching ratios. We  expect that our predictions will
provide useful guidance for future investigations in $B$ decays. We look forward to even more accurate experimental 
data  from our experimental colleagues   in order to  further constrain  our theoretical results and hence, to 
 further advance  the 
determination of the CKM parameters $\rho$ and $\eta$ and our   understanding of $CP$ violation  within or beyond
the Standard Model.
%
\subsubsection*{Acknowledgements}

This work was supported  in part by the Australian Research Council and the University of Adelaide.

\newpage

\newpage

%
\newpage
\clearpage
\clearpage
\clearpage
\clearpage

\clearpage
{\renewcommand\baselinestretch{0.94}
%
%
\begin{table}
\begin{center}
\begin{tabular}{ccc} \hline \hline    
   $C_{i}^{\prime}$          & $q^{2}/m_{b}^{2}=0.3$   & $q^{2}/m_{b}^{2}=0.5$ \\
\hline
\hline
$C_{1}^{\prime} $   &   $-0.3125$                                     &$-0.3125 $  \\
$C_{2}^{\prime} $   & $+1.1502$                                        &$+1.1502$    \\
\hline
$C_{3}^{\prime} $   & $+2.433 \times 10^{-2} + 1.543 \times 10^{-3}i$  &$+2.120 \times 10^{-2} + 2.174 
\times 10^{-3}i$\\
$C_{4}^{\prime} $   & $-5.808 \times 10^{-2} -4.628 \times 10^{-3}i$  &$-4.869 \times 10^{-2} -1.552 
\times 10^{-2}i$\\
$C_{5}^{\prime} $   & $+1.733 \times 10^{-2}+ 1.543 \times 10^{-3}i$   &$+1.420 \times 10^{-2} + 5.174 
\times 10^{-3}i$\\
$C_{6}^{\prime} $   & $-6.668 \times 10^{-2}- 4.628 \times 10^{-3}i$  &$-5.729  \times 10^{-2}- 1.552 
\times 10^{-2}i$\\
\hline
$C_{7}^{\prime} $   & $-1.435 \times 10^{-4} -2.963 \times 10^{-5}i$  &$-8.340 \times 10^{-5} -9.938 
\times 10^{-5}i$\\
$C_{8}^{\prime} $   & $+3.839 \times 10^{-4}$                          &$ +3.839 \times 10^{-4} $\\
$C_{9}^{\prime} $   & $-1.023 \times 10^{-2} -2.963 \times 10^{-5}i$  &$-1.017 \times 10^{-2} -9.938 
\times 10^{-5}i$\\
$C_{10}^{\prime} $  & $+1.959 \times 10^{-3}$                          &$+1.959 \times 10^{-3}$\\
\hline
\hline
\end{tabular}
\end{center}
\caption{Effective Wilson coefficients for  the  tree operators, electroweak and QCD penguin operators
 (see Refs~\cite{ref11, ref12}). }
\label{tab2}
\end{table}
%
%
%
\begin{table}
\begin{center}
\begin{tabular}{ccccccc} \hline \hline    
         &      $h_{A_{0}}$     & $h_{1}$    &   $m_{A_{0}}$  &   $m_{1}$  & $d_{0}(d_{1})$   &   $b_{0}(b_{1})$  \\ 
\hline
\hline
 model $(1)$   & 0.280 & 0.290  & 5.27 & 5.32 &              &               \\
\hline
 model $(2)$   & 0.340 & 0.625  & 5.27 & 5.32 &              &               \\
\hline
 model $(3)$   & 0.280 & 0.290  & 5.27 & 5.32 &              &               \\ 
\hline
 model $(4)$   & 0.340 & 0.625  & 5.27 & 5.32 &              &               \\ 
\hline
 model $(5)$   & 0.372 & 0.305  &      &      & 1.400(0.266) & 0.437(-0.752) \\ 
\hline
\hline
\end{tabular}
\end{center}
\caption{Form factor values for  $B \rightarrow \rho$ and   $B \rightarrow\pi$ at $q^{2}=0$. }
\label{tab5}
\end{table}
%
%
%
%
\begin{table}
\begin{center}
\begin{tabular}{ccc} \hline \hline    
                      & $N_{cmin}^{eff}=1.09(1.11)$   & $N_{cmax}^{eff}=1.68(1.80)$ \\
\hline
\hline
model $(1)$                                                                         \\
\hline
\hline
$ \rho_{max},\eta_{max}$   &        -55(-41)    &      -65(-51)                   \\
$ \rho_{min},\eta_{min}$   &        -72(-55)    &      -80(-65)                   \\
\hline                                  
\hline
model $(2)$                                                                        \\               
\hline
\hline
$ \rho_{max},\eta_{max}$   &        -63(-48)    &      -71(-56)                   \\
$ \rho_{min},\eta_{min}$   &        -78(-62)    &      -84(-69)                   \\
\hline  
\hline  
model $(3)$                                                                         \\
\hline
\hline
$ \rho_{max},\eta_{max}$   &        -56(-41)    &      -65(-51)                   \\ 
$ \rho_{min},\eta_{min}$   &        -72(-55)    &      -80(-69)                   \\
\hline  
\hline 
model $(4)$                                                                         \\
\hline
\hline
$ \rho_{max},\eta_{max}$   &        -64(-48)    &      -71(-57)                   \\ 
$ \rho_{min},\eta_{min}$   &        -79(-62)    &      -84(-69)                   \\ 
\hline  
\hline
model $(5)$                                                                         \\
\hline
\hline
$ \rho_{max},\eta_{max}$   &        -51(-38)    &      -58(-44)                   \\
$ \rho_{min},\eta_{min}$   &        -63(-51)    &      -72(-60)                   \\
\hline
\hline
\end{tabular}
\end{center}
\caption{Maximum $CP$ violating asymmetry $a_{max}(\%)$ for $\bar{B}^{0} \rightarrow \pi^{+} \pi^{-} \pi^{0}$, 
for all models, limiting (upper and lower) values of the  CKM matrix elements, and  $k^{2}/m_{b}^{2}=0.3(0.5)$. }
\label{tab7}
\end{table}
%
%
%
\begin{table}[p]
\begin{center}
\begin{tabular}{cccc} \hline \hline    
                       &             CLEO             &        BABAR               &  BELLE            \\ 
\hline
\hline
$\rho^{0} \pi^{\pm}$   &  ${10.4^{+3.3}_{-3.4} \pm 2.1}^{\star}$ &   ${24 \pm 8 \pm 3}^{\Box}$  $(\leq 39)^{\P}$ 
& ${8.0^{+2.3 +0.7}_{-2.0 -0.7}}^{\star}$   $(\leq 28.8)^{\P}$ \\
\hline
$\rho^{\pm} \pi^{0}$   & $\leq {43}^{\P}$                     &        $-$               &      $-$        \\
\hline
$\rho^{\pm} \pi^{\mp}$ & ${27.6^{+8.4}_{-7.4} \pm 4.2}^{\star}$  &  ${28.9 \pm 5.4 \pm 4.3}^{\star}$    & 
 ${20.8^{+6.0 +2.8}_{-6.3 -3.1}}^{\star}$   $(\leq 35.7)^{\P}$  \\ 
\hline
$\rho^{0} \pi^{0}$  &  ${1.6^{+2.0}_{-1.4} \pm 0.8}^{\bullet}$ $(\leq {5.5})^{\P} $           &   
    $\leq {10.6}^{\P} $    &         $\leq 5.3^{\P}$                \\ 
\hline
$\frac{BR(\rho^{\pm} \pi^{\mp})}{BR( \rho^{0} \pi^{\pm})}  $ & $2.65 \pm 1.9 $   &   $1.20 \pm 0.79 $  &   
$2.60 \pm 1.31$  \\
\hline
$\omega \pi^{\pm}$ & ${11.3^{3.3}_{-2.9} \pm 1.4}^{\star}$ & ${6.6^{2.1}_{-1.8} \pm 0.7}^{\star}$ & ${4.2^{2.0}_{-1.8}
\pm 0.5}^{\star}$ \\
\hline
\hline
\end{tabular}
\end{center}
\caption{The branching ratios measured by CLEO, BABAR and BELLE factories for $B$ decays into $\rho \pi$ in unit of
$10^{-6}$ (see Ref in text). 
 ${\rm Experimental \; data}^{\star}$, ${\rm preliminary \; results}^{\Box}$, ${\rm fit}^{\bullet}$ and ${\rm upper \;
 limit}^{\P}$.} 
\label{tab6}
\end{table}
%
%
%
\newpage

%
%
\begin{table}
\begin{center}
\begin{tabular}{cc} \hline \hline   
    $B \rightarrow \rho \pi$    &   $\left\{N_{c}^{eff}\right\}$ with mixing   \\
\hline \hline 
model $(2)$               & 1.09;1.63(1.12;1.77)                 \\
model $(4)$               & 1.10;1.68(1.11;1.80)                \\   

\hline
\hline
maximum range           & 1.09;1.68(1.11;1.80)                 \\
minimum range           & 1.10;1.63(1.12;1.77)                     \\
\hline
\hline
\end{tabular}
\end{center}
\caption{Best range of  $N_{c}^{eff}$ determined for 
$k^{2}/m_{b}^{2}=0.3(0.5)$ and for all $ B \rightarrow \rho \pi$ decays.}
\label{tab15}
\end{table}}
%
%
\end{document}